\input harvmac
\input epsf
%
\rightline{EFI-96-47, WIS/96/46/Dec-Ph}
\Title{
\rightline{hep-th/9612102}
}
{\vbox{\centerline{M-branes and N=2 Strings}}}
\medskip
\centerline{\it David Kutasov\footnote{*}
{On leave of absence
from Department of Physics, University of Chicago,
5640 S. Ellis Ave., Chicago, IL 60637, USA.} }
\smallskip
\centerline{Department of Physics of Elementary Particles}
\centerline{Weizmann Institute of Science}
\centerline{Rehovot, 76100, Israel}
\vskip .2in

\centerline{\it Emil Martinec\footnote{**}{Supported 
in part by Dept. of Energy grant DE-FG02-90ER-40560.  } 
}
\smallskip
\centerline{Enrico Fermi Inst. and Dept. of Physics}
\centerline{University of Chicago}
\centerline{5640 S. Ellis Ave., Chicago, IL 60637, USA}

\vglue .3cm
\bigskip
 
\noindent
The string field theory of N=(2,1) heterotic strings describes
a set of self-dual Yang-Mills fields coupled to
self-dual gravity in 2+2 dimensions. 
We show that the exact classical action for this field theory
is a certain complexification
of the Green-Schwarz/Dirac-Born-Infeld string action, closely 
related to the four dimensional Wess-Zumino action describing 
self-dual gauge fields. This action describes the world-volume 
of a 2+2d ``M-brane'', which gives rise upon different null
reductions to critical strings and membranes.
We discuss a number of further properties of N=2 heterotic strings,
such as the geometry of null reduction, general features of a  
covariant formulation, and possible relations to BPS and GKM algebras.

\Date{12/96}

\def\journal#1&#2(#3){\unskip, \sl #1\ \bf #2 \rm(19#3) }
\def\andjournal#1&#2(#3){\sl #1~\bf #2 \rm (19#3) }

\def\ie{{\it i.e.}}
\def\eg{{\it e.g.}}
\def\cf{{\it c.f.}}
\def\etal{{\it et.al.}}
\def\etc{{\it etc.}}

\def\sst{\scriptscriptstyle}

\def\frac#1#2{{#1\over#2}}
\def\coeff#1#2{{\textstyle{#1\over #2}}}
\def\half{\frac12}
\def\hf{{\textstyle\half}}

\def\vev#1{\langle#1\rangle}
\def\d{\partial}

\def\inbar{\,\vrule height1.5ex width.4pt depth0pt}
\def\IC{\relax\hbox{$\inbar\kern-.3em{\rm C}$}}
\def\IR{\relax{\rm I\kern-.18em R}}
\def\IP{\relax{\rm I\kern-.18em P}}

%
%
\def\np#1#2#3{Nucl. Phys. {\bf B#1} (#2) #3}
\def\pl#1#2#3{Phys. Lett. {\bf #1B} (#2) #3}

\def\prl#1#2#3{Phys. Rev. Lett. {\bf #1} (#2) #3}

\def\prd#1#2#3{Phys. Rev. {\bf D#1} (#2) #3}

\def\cmp#1#2#3{Comm. Math. Phys. {\bf #1} (#2) #3}
\def\cqg#1#2#3{Class. Quant. Grav. {\bf #1} (#2) #3}
\def\mpl#1#2#3{Mod. Phys. Lett. {\bf #1} (#2) #3}

\catcode`\@=11
\def\slash#1{\mathord{\mathpalette\c@ncel{#1}}}
\overfullrule=0pt

\def\EE{{\cal E}}
\def\FF{{\cal F}}

\def\LL{{\cal L}}
\def\MM{{\cal M}}

\def\lam{\lambda}
\def\eps{\epsilon}

\def\underrel#1\over#2{\mathrel{\mathop{\kern\z@#1}\limits_{#2}}}

\catcode`\@=12


%

\def\vev#1{\left\langle #1 \right\rangle}
\def\det{{\rm det}}

\def\det{{\rm det}}
\def\exp{{\rm exp}}


\def\psibar{{\bar\psi}}

\def\alphabar{{\bar\alpha}}
\def\betabar{{\bar\beta}}

\def\thetabar{{\bar\theta}}
\def\abar{{\bar a}}
\def\bbar{{\bar b}}
\def\ibar{{\bar i}}
\def\jbar{{\bar j}}
\def\kbar{{\bar k}}
\def\lbar{{\bar \ell}}
\def\mbar{{\bar m}}
\def\dbar{{\bar \d}}
\def\s{{\bf S}}
\def\ij{{i\bar j}}
\def\kahler{{K\"ahler}}
\def\ferm{{\vartheta}}
\def\fermbar{{\bar\vartheta}}


\newsec{Introduction}

All known (weakly coupled) string theories, as well as eleven-dimensional
supergravity, appear to be different manifestations of a unique theory,
compactified on different manifolds and studied in various extreme
limits in the moduli space of vacua.  
It is an outstanding problem to find a good presentation of this theory
which is valid everywhere in moduli space.  The degrees of freedom
appropriate for such a presentation are unknown.  
The problem is especially challenging since the unified theory
should have the property that:

\item{a)} in different limits it looks like strings with different
worldsheet gauge principles, or even as an eleven-dimensional theory
which doesn't have a string description at all;

\item{b)} the dilaton (or string coupling) should play the role
of a size of a compact manifold, and in particular appear on the
same footing as other geometrical data (``U-duality'').

\noindent
All this is in sharp contrast with the conventional description
of any single string vacuum, where the worldsheet gauge principle
is chosen at the outset, and the string coupling appears in
a very different way than the geometrical moduli.

\nref\km{D. Kutasov and E. Martinec, hep-th/9602049, 
Nucl. Phys. {\bf B477} (1996) 652.}%
\nref\kmo{D. Kutasov, E. Martinec and M. O'Loughlin,
hep-th/9603116, Nucl. Phys. {\bf B477} (1996) 675.}%
In two recent papers 
\refs{\km,\kmo}, it was pointed out
that N=2 heterotic strings may provide important clues regarding
the degrees of freedom appropriate for a more fundamental
formulation of the theory.  N=2 heterotic strings live in a 2+2
dimensional target space, with a null reduction restricting the 
dynamics to 1+1 or 2+1 dimensions.  The main results
of \refs{\km,\kmo} are:

\nref\green{M. Green, \np{293}{1987}{593}.}%
\item{1)} The target space dynamics of critical N=2 heterotic strings 
describes critical string worldsheets and membrane worldvolumes
in static gauge (see also \green\ for early work).

\item{2)} All types of ten-dimensional superstring theories, as well
as the eleven-dimensional supermembrane, arise in different limits
of the moduli space of N=2 strings.  One can continuously
interpolate between them by varying the moduli of the N=2 string.
Thus, N=2 strings appear to be the `building blocks' of critical
(super)string/membrane theories.

\item{3)} The 1+1 dimensional dynamics on a superstring worldsheet 
and the 2+1 dimensional worldvolume dynamics of a supermembrane 
appear to be different manifestations of the 2+2 dimensional
worldvolume dynamics of an `M-brane' propagating in a 10+2
dimensional spacetime.  The null reduction yields different
string worldsheets and membrane worldvolumes.

\noindent
The 2+2/10+2 dimensional perspective sheds new light on string dynamics.
Scalars on the string worldsheet of the usual formalism describe
the embedding of the string in spacetime; on the M-brane,
they are replaced by self-dual
gauge fields.  The abelian worldsheet gauge field familiar
from D-strings is replaced by a self-dual four-dimensional metric.
The principle of reparametrization invariance and the residual
conformal invariance on the string worldsheet are replaced by
a not yet fully understood symmetry principle, apparently related 
to four-dimensional self-duality.
The appearance
of extra dimensions allows relations among M-theory
vacua that are hidden in other approaches. Since some of these relations
are due to strong/weak coupling duality, the hope is that
the study of M-branes will eventually lead to a non-perturbative 
formulation of string theory.

The 2+1 dimensional vacua of the theory describe membrane-like objects;
because these objects appear in the theory on the same footing
as strings and are embedded in the underlying 2+2 self-dual theory,
it is a strong possibility that these membranes can be quantized, leading
to a definition of `M-theory' (see 
\ref\bfss{T. Banks, W. Fischler, S.  Shenker and L.
Susskind, hep-th/9610043.} for another suggestion in this regard).

There are two possible ways of viewing the role of the N=2 string in 
our construction, both of which are useful to keep in mind.
One is to regard the N=2 string as merely
a tool to probe the 2+2 dimensional
dynamics on the M-brane worldvolume, which in this view is
`fundamental'.  If one adopts this point of view, the task is
to extract as much information as possible about the 2+2 field theory;
then study its quantization as a field theory (perhaps by
using heterotic N=2 string field theory as a tool). 
Alternatively, one may take seriously the idea that
N=2 heterotic strings are the `constituents' out of which critical
strings and membranes are made\foot{More precisely, since the N=2
string so far only seems to give strings and membranes in static
gauge, a more conservative interpretation is that N=2 strings are
the quanta of the Goldstone modes; the constituents may be something
else.  A useful analogy might be two-dimensional noncritical strings,
which are the quanta of a theory whose underlying constituents
are free fermions.}, and take seriously all physical
fluctuations of N=2 strings.
If one takes this stance, the problem of quantization of the unified
theory turns into the problem of second (or third) quantization
of the heterotic N=2 string. 

At the current level of understanding, the second point of view appears
to be more viable for two reasons:

\item{a)} As we'll discuss below,
the quantization of the 2+2d M-brane worldvolume theory 
implied by N=2 string field theory appears to
be highly unconventional. For example, compactification 
of the 2+2d worldvolume reveals a rich spectrum of new degrees
of freedom (arising from
mixed momentum/winding excitations of the
N=2 heterotic string), unseen in the infinite radius limit,
and absent in the 2+2d worldvolume theory. These degrees of
freedom appear to be needed for a ``microscopic'' understanding of
string duality; in fact, in the recent work of
Dijkgraaf, Verlinde, and Verlinde
\ref\dvv{R. Dijkgraaf, E. Verlinde and
H. Verlinde, hep-th/9603126, hep-th/9604055, hep-th/9607026.}, 
a similar construction
of an `underlying string' was claimed to shed light on the
microscopic degrees of freedom needed for heterotic string S-duality
in D=4.  

\item{b)} In addition to the class of vacua describing 
supersymmetric 2+2 branes living in 10+2 dimensions,
we'll discuss below vacua of heterotic N=2 
strings describing other kinds of critical strings in various
dimensions (26d bosonic,
10d fermionic and various N=2 strings in target space). 
While the relation between the five known
ten dimensional superstring theories and
eleven dimensional supergravity can be understood 
in our framework directly on the 2+2d worldvolume
of the M-brane, the other classes of strings can only
be understood by appealing to the underlying heterotic
N=2 string.

\noindent
The main argument against viewing N=2 heterotic strings
as fundamental is that so far they have only given
particularly symmetric points in the moduli space
of string vacua, and it has proven difficult to construct
the full moduli space this way. We will comment on this
issue below.
 
In any case, the sort of `string miracles' 
occurring in the N=2 heterotic string by now seem
rather unlikely to be an extraneous coincidence,
and strongly suggest that N=2 strings must be incorporated 
into the emerging picture of dualities in the master theory.

The purpose of this paper is to report on the status
of the program of \km\ of studying M-branes using
their realization as collective excitations of N=2
strings, and in particular present new results regarding
the target space dynamics of N=2 heterotic strings.  
Heterotic geometry is described by a gravity multiplet --
metric, antisymmetric tensor, and dilaton --
coupled to gauge fields.  The geometrical setting is thus a vector
bundle $\EE$ over a manifold $\MM$.  As we describe in detail
in sections 2 and 3, in N=(2,1) heterotic geometry the critical
dimension is four, the geometry is self-dual, and $\EE$ is
an adjoint bundle.  The proposal of \km\ is to regard the 
four-manifold (of signature 2+2) as the world-volume of 
an extended object embedded in the total space of the 
bundle $\EE\rightarrow\MM$ (see figure 1).

{\vbox{{\epsfxsize=2.5in
    \centerline{\epsfbox{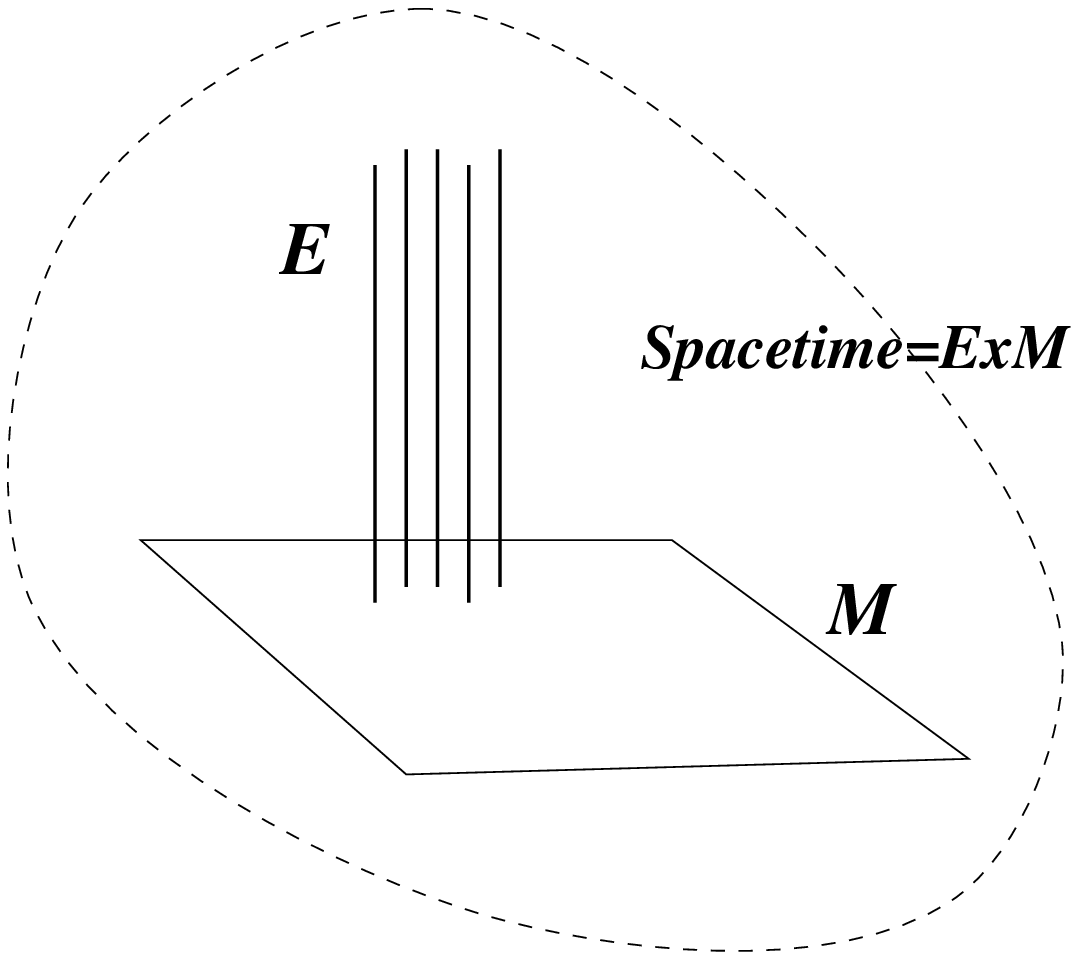}}
    {\raggedright\it \vbox{
{\bf Figure 1. }{\it M-brane geometry.  }
 }}}}
\bigskip}

\noindent
Consistency of the (2,1) string requires a null reduction to
be imposed on the geometry.  Strings are obtained when the null
vector field lies entirely in the base $\MM$; membranes when its
timelike part lies in the base and its spacelike part lies in the 
fiber $\EE$.
The ripples on the brane are described by a metric $e^\phi$ on the fibers,
which specifies the self-dual gauge connection
by the usual Yang ansatz.  Our main result is
that the dynamics of this arrangement is described by the 
geometrical action
\eqn\mainresult{
  S_{2+2}=\int d^4x\sqrt{\det[g_\ij + \d_i\phi^a\d_\jbar\phi^a]}
}
where $g_{i\bar j}$ is the self-dual metric with torsion on
the base \MM,
and the scalars $\phi^a$ parametrize the self-dual gauge connection
in the case where the gauge group is abelian (the nonabelian extension
is also described).

The plan of the paper is as follows.
In section 2 we briefly review the technology of 
constructing N=2 heterotic string vacua. We emphasize the
fact that the 1+1 dimensional vacua of N=2 heterotic strings
seem always to describe in their target spaces critical
string worldsheets in static gauge, and give a number
of new examples of this phenomenon. There appears to be
a close relation between the properties of the target
worldsheet and the underlying N=2 one. We then discuss
the 2+2 dimensional target space geometry of N=2 
heterotic strings, which underlies the dynamics of the 1+1
and 2+1 dimensional string/membrane vacua, and describes
a non-linear self-dual system of gauge fields coupled to 
gravity. We also describe the striking change the theory
undergoes upon compactification, and point out the relations to
the physics of BPS states and string duality.

Section 3 constructs the classical 2+2d dynamics
on the worldvolume of the M-brane
(for earlier work, see \ref\lawrence{A. Lawrence,
hep-th/9605223.}). 
It is known that scattering amplitudes
with more than three external particles vanish in N=2 string theory.
Nevertheless, the target space action \mainresult\ is 
non-polynomial; we determine this action exactly
in the classical limit of small N=2 string coupling constant
(large M-brane tension).
This is achieved by a combination
of string S-matrix and beta-function techniques, which
allow one to argue in this case that the one loop sigma
model beta function is {\it exact}.

Section 4 contains discussions of a number of open issues, such
as the geometrical role of the null reduction and its possible
relation to conformal symmetry.  We outline the ingredients 
required for a covariant formulation of the dynamics, and give
a brief treatment of the
generalizations necessary to include winding N=2 strings.
The latter lead to intriguing connections to BPS algebras
\ref\harvmooretwo{J. Harvey and G. Moore, hep-th/9510182, 
hep-th/9609017.}
and GKM algebras such as the monster Lie algebra.
At various points,
we comment on the possible relation of other recent work
to ours, including F-theory 
\ref\vafa{C. Vafa, hep-th/9602022; \np{469}{1996}{403}.}
and matrix models of M-theory \bfss.
Two appendices discuss the different
null reductions, and some of the reasons to expect the quantization 
of the 2+2d M-brane worldvolume theories to be subtle.

\newsec{Some properties of N=2 heterotic strings.}

\nref\ademollo{
M. Ademollo, L. Brink, A. D'Adda, R. D'Auria, E. Napolitano, S. Sciuto, 
E. Del Giudice, P. Di Vecchia, S. Ferrara, F. Gliozzi, 
R. Musto, R. Pettorino, J. Schwarz,
\np{111}{1976}{77}.}%
\nref\marcus{N. Marcus,
talk at the Rome String Theory Workshop (1992); hep-th/9211059.}%
\nref\ovone{H. Ooguri and C. Vafa, \np{361}{1991}{469}.}%
Since their discovery twenty years ago \ademollo, 
N=2 strings have been
among the most enigmatic types of string theories. As we will 
review below (see \eg\ \marcus, \ovone\ for more on N=2 strings),
unlike other critical string theories, critical N=2 strings
live in a low-dimensional target space (two, three, or four 
dimensional depending on the particular model and background); 
they describe a finite number
of field theoretic degrees of freedom in target space, whose 
dynamics has yet to be fully elucidated.
We will focus on heterotic N=2
strings \ref\ovtwo{H. Ooguri and C. Vafa, \np{367}{1991}{83}.}; 
our interest in these theories stems from the observation
of \refs{\km,\kmo} mentioned above, that the target space dynamics of
N=2 heterotic strings generates all known 
classes of M-theory vacua,
and seems to offer an interesting perspective on the problem
of finding a unified description of all such vacua.

The target space dynamics of N=2 heterotic strings appears
to replace conformal invariance on the string worldsheet by
self-duality on a 2+2 dimensional worldvolume (the target
space of the N=2 string). The natural objects defined on the
worldsheet -- scalar fields describing the embedding of the 
worldsheet in space-time and the Born-Infeld gauge field
-- are replaced by self-dual gauge fields
and metric on the worldvolume, respectively. 
One of our primary
tasks in this paper will be to study the dynamics of this
non-linear 2+2d self-dual system.

Upon compactification on a circle one finds a qualitative change
in the target space physics. The finite number of target
space fields is replaced by a tower of states of arbitrarily
high mass. The structure is very reminiscent of the construction
of certain BPS states in string theory, and indeed we'll see
below that there are close parallels with recent discussions
of the physics of BPS states in type II/heterotic string theories.
The density of states with fixed momentum/winding around the 
circle grows as $\rho(E)\sim E^\alpha\exp(\beta\sqrt E)$
(with $\alpha$, $\beta$ certain known constants), a fact that
ensures that the dynamics is still relatively simple.  
The compactified N=2 heterotic string exhibits important connections
between worldsheet and space-time physics, and its better
understanding seems important for understanding string duality.

In this section we will summarize some of
the known facts about N=2 heterotic
strings\foot{See \km, \kmo\ for additional discussion and references
to earlier work.}, as well as make a few new points.
We start with a review of the worldsheet construction of N=2 
heterotic strings, followed by a discussion of the target space
dynamics, and then an outline of the compactified case.

\subsec{N=2 heterotic worldsheets.}

Like the more familiar N=(1,0) heterotic string, N=2
heterotic strings have the property that left and right
moving modes on the worldsheet couple to different 
(super) gravity theories. The right moving modes 
always couple to N=2 worldsheet supergravity. For
N=(2,0) heterotic strings, the left movers describe 
a critical bosonic string, while for N=(2,1) strings
the left movers couple to N=1 supergravity. The critical
central charge for the right movers is $\bar c=6$, which
can be realized by four right moving superfields
$(x^\mu, \bar\psi^\mu)$, $\mu=0,1,2,3$ with signature
$(-,-,+,+)$ living in a flat space-time $\IR^{2,2}$ 
with metric $\eta_{\mu\nu}$ (we
will discuss the interesting case of compact $x^\mu$ later).
The peculiar signature of space-time is dictated by the necessity
of choosing a complex structure, which we will denote
by $I_{\mu\nu}$. The complex structure satisfies:
\eqn\iss{I_{\mu\nu}=-I_{\nu\mu};\;\;\; I_{\mu\nu}I^{\nu\lambda}=
\eta_\mu^\lambda.} 
The choice of complex structure breaks the Lorentz 
symmetry\foot{One can think of the breaking
as spontaneous and interpret $I_{\mu\nu}$ as an expectation
value of a target space field.}
from $O(2,2)$ to $U(1,1)$. The right moving N=2 superconformal
gauge algebra, which consists of the stress tensor $\bar T$,
two superconformal generators $\bar G^\pm$, and a $U(1)$ current
$\bar J$, is represented on $x^\mu, \bar\psi^\mu$ as:
\eqn\ntwo{\eqalign{
\bar T=&~-\half\bar\partial x\cdot 
\bar\partial x-\half\bar\psi\cdot\bar\partial
\bar\psi\cr
\bar G^\pm=&~(\eta_{\mu\nu}\pm I_{\mu\nu})\bar\psi^\mu\bar\partial x^\nu\cr
\bar J=&~\half I_{\mu\nu}\bar\psi^\mu\bar\psi^\nu.\cr}
}
The four non-compact scalar fields $x^\mu$ are shared by the left
and right moving sectors. To describe the rest of the left
moving structure it turns out to be essential \ovtwo\ to
enlarge the left moving gauge principle by a $U(1)$ gauge symmetry.
We next describe the resulting structure for the two kinds of
N=2 heterotic strings, starting with the N=(2,0) one.

\bigskip

\noindent{\it N=(2,0) heterotic strings.}

\smallskip

The conformal gauge generators for the left movers are
the stress tensor $T(z)$, and the $U(1)$ current $J(z)$.
The critical dimension is $c=28$, which can be realized
in terms of twenty-eight 
scalar fields $x^a$, $a=0,1, \cdots, 27$, with
signature $(-,-, (+)^{26})$. The generators
\eqn\tj{\eqalign{
T(z)=&~-{1\over2}\partial_z x^a\partial_z x^a\cr
J(z)=&~v^a\partial_z x^a\cr
}}
involve a choice of a preferred null vector $v^2=0$. Physical 
states satisfy
\eqn\phst{J_n|{\rm phys}\rangle=0;\;\;n\geq0}
which effectively decouples two of the twenty-eight dimensions,
giving back the familiar critical dimension of bosonic
strings, twenty-six, and the usual Minkowski signature.

What kinds of vacua of N=(2,0) strings are there?  Keeping
$(x^0, \cdots, x^3)$ non-compact, modular invariance requires
the chiral scalars $(x^4, \cdots, x^{27})$ to live on an even,
self-dual Euclidean torus, one of the twenty-four Niemeier tori. 
The worldsheet partition sums for the resulting theories
appear in Appendix A. Additional
possibilities involving orbifolds and twisting of the N=2 algebra
will be mentioned below. 

Depending on the choice of the null vector $v$ \tj, the 
theory can be thought of as living in 1+1 or 2+1d,
corresponding, respectively, to $v$ pointing in the first
four directions or having its timelike component lie in the 0-1
plane and its spacelike component in the $(x^4, 
\cdots, x^{27})$ hyperplane\foot{One might have thought
that a third possibility is to orient the null vector
entirely along the
Niemeier torus, with  $v$ a complex null
vector, such as $v=\hat x^4+i\hat x^5$. However, as
we show in Appendix A, this option is inconsistent, leading
to a non-modular invariant theory.}.
The spectrum of physical states for the two cases is:

\medskip
\noindent
\underbar{\sl 1+1d}: A massless scalar for every dimension one
operator in the meromorphic c=24 conformal field theory
(CFT) describing $(x^4, \cdots, x^{27})$. These states
transform in the adjoint representation of a rank twenty-four 
simply laced group $H$ characterizing the Niemeier lattice (\eg\
$U(1)^{24}$, $E_8^3$, $SU(25)$, etc). Their vertex operators
are
\eqn\vbos{V^a_\phi=J^a(z)\int d^2\bar\theta\; e^{ik\cdot x};\;\;\;
a=1,\cdots, {\rm dim} \; H}
where the $\bar\theta$ integrals schematically denote the
raising operators $\bar G^\pm_{-{1\over2}}$. 
The currents $J^a$ are given by the standard construction; 
\eg\ the Cartan subalgebra (CSA) generators correspond
to $J^a=\partial x^a$.

\medskip
\noindent
\underbar{\it 2+1d}: Take for concreteness
\eqn\vex{v=(1,0,0,0,1,0,0,\cdots).}
The theory now lives on the 2+1d space-time parametrized
by $(x^1, x^2, x^3)$. On this space-time we find a massless
gauge field with vertex operator
\eqn\gf{V_K=\xi_\mu\partial x^\mu\int d^2\bar\theta\; e^{ik\cdot x}}
with $\xi\cdot k=0$, $\xi\sim \xi+\epsilon k$; $k=(k^1, k^2, k^3)$,
$k^2=0$. In addition we find scalars in the adjoint representation
of a rank 23 group whose CSA is spanned by $J^a=\partial x^a$,
$a=5,\cdots, 27$. States carrying non-zero charge $Q_4$ under
$J^4=\partial x^4$ become tachyonic, since
\eqn\dispr{E^2\equiv k_1^2=k_2^2+k_3^2-k_0^2}
and at the same time, the condition \phst\ relates $k_0$ to the
charge $Q_4$: $k_0=Q_4$. The tachyonic dispersion
relation \dispr\ should lead to destabilization of the perturbative
vacuum (see also Appendix A), and it would be 
interesting to understand this in more detail. 

Note also that the vacua of N=(2,0) strings built on Niemeier
lattices described above are not the only ones that exist. For the
case of non-compact $(x^0, \cdots, x^3)$, vacua of the N=(2,0)
string are in one to one correspondence with modular invariant
meromorphic CFT's with c=24; there are quite a few of these (see
\eg\
\ref\schell{B. Schellekens, hep-th/9205072, 
\cmp{153}{1993}{159}.}).

\bigskip

\noindent{\it N=(2,1) heterotic strings.}

\smallskip

Here, we find a left moving N=1 superconformal gauge algebra
acting on the $\hat c=12$ system consisting of the superfields
$(x^a, \psi^a)$, $a=0, \cdots, 11$, where, again, $(x^0, \cdots,
x^3)$ are non-compact, while the rest of the $\{x^a\}$ are chiral, 
compact scalars.  The superstress tensor can be written as:
\eqn\none{\eqalign{
T=&~-\half(\partial x)^2-\half\psi\partial\psi\cr
G=&~\psi\partial x\cr
}}
and again it is convenient to add a gauged $U(1)$ supercurrent
of the form:
\eqn\nullu{\eqalign{
J=&~v^a\partial x^a\cr
\Psi=&~v^a\psi^a\cr
}}
As in \phst, $J$, $\Psi$ must annihilate physical states,
and depending on the direction of $v$ the theory lives
in 1+1 or 2+1d. Vacua of the uncompactified N=(2,1)
string are in one to one correspondence with meromorphic
Superconformal Field Theories (SCFT's) with $\hat c=8$
($c=12$), describing the ``compact'' chiral sector 
$(x^a, \psi^a)$, $a=4, \cdots, 11$. We are not aware of 
a classification of such SCFT's, but it is easy to construct
examples. 

One example is obtained by compactifying 
the $\{x^a\}$, $a=4, \cdots, 11$ on the $E_8$ torus. 
This gives rise to the partition sum\foot{Actually, this 
partition sum is appropriate to the 1+1d vacuum of the theory.
See Appendix A for the modification needed for the
2+1d case.}
\eqn\ztwo{Z(\tau)={1\over4}
\left[
\left({\theta_3\over\eta}\right)^4-
\left({\theta_4\over\eta}\right)^4-
\left({\theta_2\over\eta}\right)^4
\right]
\left[
\left({\theta_3\over\eta}\right)^8+
\left({\theta_4\over\eta}\right)^8+
\left({\theta_2\over\eta}\right)^8
\right]\ .}
with the two factors in brackets arising from the contributions
of $\{\psi^a\}$ and $\{x^a\}$, respectively.
The spectrum of the 1+1d theory includes in this case eight
bosonic states with vertex operators
\eqn\vb{
  V^a_\phi=\int d\theta\int d^2\bar\theta\;\psi^a e^{ik\cdot x}\ ;
	\qquad a=4, \cdots, 11.
}
We use $\int d\theta$ and $\int d^2\thetabar$ as shorthand for the action
of the superconformal generators
$G_{-\frac12}$, ${\bar G}_{-\frac12}^\pm$.
In the 2+1d vacuum we find seven states of the form
\vb, respectively, however in addition to those and in 
analogy to \gf, a massless gauge field appears:
\eqn\gfone{V_K=\int d\theta\int d^2\bar \theta\; \xi_\mu\psi^\mu e^{ik
\cdot x}} with the standard gauge invariance, $\xi\sim \xi+\epsilon k$,
$\xi\cdot k=0$.
Note, in particular, that the total number of physical bosonic
degrees of freedom is always eight. 

A quick glance at the partition sum \ztwo\ reveals a new feature
of the N=(2,1) string -- the appearance of space-time fermions, with
vertex operators
\eqn\vf{
  V_\ferm=u^\alpha(k)\Sigma S_\alpha\int d^2\bar\theta\; e^{ik\cdot x}
}
where $S_\alpha$ is the dimension 3/4, 32 component spin field for 
the twelve fermions $\psi^a$, and $\Sigma=\exp(-\phi/2+\rho/2)$
is a dimension 1/4 spin field for the superconformal and super $U(1)$
ghosts (see \kmo\ for details). The null reduction \nullu\ eliminates
half of the components of $V_\ferm$, while the Dirac equation eliminates
half of the remaining ones. We are left with eight fermionic 
degrees of freedom,
equal to the number of bosonic ones. 
In fact the system is supersymmetric \refs{\km, \kmo}; 
the supercharges
$Q_\alpha$ are given by
\eqn\supch{Q_\alpha=\oint dz\; \Sigma S_\alpha}
where, again, only half of the 32 components are physical
due to the null constraints \nullu. 
The unbroken supercharges transform in the
16 of the Spin(10) subalgebra of Spin(12) preserving the
form of \nullu. 

Another type of vacuum is obtained by fermionizing the eight scalars
$x^a$, $a=4, \cdots, 11$, and realizing the meromorphic 
$\hat c=8$ SCFT in terms of the resulting twenty-four fermions. 
A diagonal sum over spin structures of all twenty-four fermions 
\eqn\zone{
  Z(\tau)={1\over2}\left[
	\left({\theta_3\over\eta}\right)^{12}-
	\left({\theta_4\over\eta}\right)^{12}-
	\left({\theta_2\over\eta}\right)^{12}
	\right]
}
leads \ref\pierce{D. Pierce, hep-th/9511160.},
\km\ to a theory with only bosonic physical states,
\eqn\vvv{V^A_\phi=\int d\theta\int d^2\bar\theta\; \psi^A
e^{ik\cdot x}}
transforming in the adjoint of a twenty-four 
dimensional group $H$.\foot{Note
that in this case $H$ need not be simply laced.} 
The superconformal generator for the twenty-four
fermions is given in terms
of the structure constants of $H$:
\eqn\gff{G=f_{ABC}\psi^A\psi^B\psi^C;\;\;\; A, B, C=1, \cdots, 24.}
Note that the form \gff\ implies that the group $H$ can not 
contain any $U(1)$ factors.

An interesting question for all the vacua of N=(2,0), (2,1)
strings described above is what target space dynamics they
describe. Before turning to that, we would like to comment
on the issue of $U(1)$ instantons and $\theta$ vacua for N=2 strings.

\bigskip

\noindent{\it Worldsheet instantons and $\theta$ vacua.}

\smallskip

N=2 heterotic strings have in fact {\it two} independent
$U(1)$ gauge fields coupling to the left and right movers.
It is usually stated that the two are components of a single 
gauge field $A_\mu=(A, \bar A)$. However, it is possible
to perform separate gauge transformations on $A$ and $\bar A$:
\eqn\gainv{A\to A+\partial \epsilon;\;\; \bar A\to \bar A+\bar\partial
\bar\epsilon.}
This is because the gauge fields $A, \bar A$ couple to currents
$\bar J$, $J$, which are separately anomaly free. 
One can therefore try to add $\theta$ terms corresponding to
both gauge fields to the Lagrangian:
\eqn\lth{\LL=\theta F+\bar\theta\bar F.}
In fact, since the left moving $U(1)$ current $J$ \tj, \nullu\ is
non-compact, physics is independent of $\bar\theta$ in \lth\foot{
For N=(2,2) strings, both $J$ and $\bar J$ are compact and 
amplitudes depend on both $\theta$ angles.}.

A natural question regards the dependence of the physics
on $\theta$, \ie\ how do $U(1)$ instantons modify the dynamics?
In the matter sector, the effect of an instanton is to induce 
spectral flow by one unit of U(1) charge; for instance this shifts
$\bar J=\eta_{\ij}\psibar^i\psibar^\jbar$ to 
${\bar J}^+=\eps_{ij}\psibar^i\psibar^j$.
It is not difficult to show that the entire effect of instantons
can be incorporated as a modification of $I_{\mu\nu}$. 
This was observed in the context of N=(2,2) strings (and used
to great effect) in 
\nref\berkvafa{N. Berkovits and C. Vafa, 
hep-th/9407190; \np{433}{1995}{123}.}%
\nref\ovthree{H. Ooguri and C. Vafa, hep-th/9505183;
\np{451}{1995}{121}.}%
\refs{\berkvafa, \ovthree}.
In complex coordinates we can take:
\eqn\iij{\eqalign{
	I_{i\bar j}=&~a\eta_{i\bar j}\cr
	I_{ij}=&~b\epsilon_{ij}\cr
	I_{\bar i\bar j}=&~b^*\epsilon_{\bar i\bar j}\cr
}}
with $a$ real and $b$ complex, and $a^2+|b|^2=1$ because of \iss. 
One can then think of $\theta$ as the phase of $b$.

The choice of $I_{\mu\nu}$ breaks O(2,2) symmetry of the target space;
if there were no other background data violating Lorentz
invariance, all choices would be equivalent.  However, N=2 heterotic
string backgrounds also involve a choice of null vector $v$.
Thus the physical data are the relative orientation of $v$ 
and $I_{\mu\nu}$.  For instance, in the reduction to 1+1d, we can
choose coordinates so that the null reduction is $x^i=x^\ibar$,
and the complex structure is arbitrary as in \iij.
This is technically simpler than fixing, say, $b=0$ in
the choice of $I_{\mu\nu}$, and working with an arbitrary
null reduction, and it is what we'll do below.

\subsec{Space-time features of N=2 heterotic strings.}

As discussed in the previous section, N=2 heterotic strings
describe in their two or three dimensional 
target space a field theory
coupled to gravity.  What is that theory?

Consider first the two dimensional case. We know of a consistent
theory of matter coupled to gravity in 2d -- worldsheet
string theory. Thus one may ask what is the relation between
the target space dynamics of N=2 heterotic strings in their 2d
vacua and worldsheet string theory.
In \refs{\km, \kmo} evidence has been presented for the validity
of the following conjecture:

\item{}{\it The two dimensional target space dynamics of critical N=2
heterotic strings always describes critical string worldsheet
dynamics in physical gauge. }

\noindent
It should be emphasized that there is at present no conceptual
understanding of why this conjecture should hold; the fact
that it does seems even in hindsight surprising. 
Since, as argued in \refs{\km, \kmo}, this presentation of string
worlsheets appears to be closely related to string duality,
it is important to verify and further understand this issue.

We next turn to a few examples of N=2 string vacua and the target
space strings they describe.

\medskip

\noindent{{\sl Example 1}: N=(2,0) strings on Niemeier tori.}

\smallskip

In the previous subsection we saw that the physical excitations 
of an N=(2,0) string on a particular Niemeier torus characterized by 
a simply-laced group $H$ of rank twenty-four
are scalars $g$ in the adjoint of $H$.
What is the dynamics of these scalars?  We will see that it is
governed by the WZW lagrangian (coupled to gravity); 
for the above conjecture to hold, the level should be $k=1$,
since only then is the target string critical.
We should emphasize that we are
talking here about a WZW action in the 1+1d {\it target} space of the
N=2 string; there is a similar (but chiral) WZW action describing
the worldsheet degrees of freedom, and in fact there is 
the possibility of an interesting
`duality' between the two; \eg\ it is natural for
the level $k$ of the affine
Lie algebra $\hat H$ to be the same on the worldsheet
and in target space.
We should also stress that, at the moment, we have no idea why or how
the level $k$ should be fixed at this particular value
(likewise for the special value that seems appropriate for
N=(2,1) strings, example 4 below).
It could be that other values of $k$ are allowed, in which case the 
target worldsheet theory would be a noncritical string,
with nontrivial dynamics for worldsheet gravity.

It is well-known that the level one WZW models for all simply laced 
groups lead to theories with $c={\rm rank}(H)$, which 
in the present case ${\rm rank}(H)=24$ are equivalent
to twenty-four free scalars living on a torus.  Therefore, the 1+1d target
space theory of the N=(2,0) string corresponds in this case to a 
critical bosonic string compactified to two dimensions on a 
twenty-four dimensional torus
(presumably the same Niemeier torus appearing 
in the worldsheet construction).
Equivalently, one can think of these vacua as describing
critical bosonic strings compactified on a rank twenty-four 
group manifold.

The simplest example of a Niemeier torus is the Leech torus which has
no vectors of length $\sqrt2$, and thus has twenty-four
physical states 
corresponding to $H=U(1)^{24}$. These twenty-four
scalar fields are governed
by the Nambu-Goto action and describe the transverse dimensions of a 
critical string.

An interesting point concerns the N=(2,0) string coupling $\lambda$,
which determines the coupling of the target space WZW model. It is well
known that the WZW model describes an RG flow from a free UV fixed
point with central charge $c_{{\sst UV}}={\rm dim} H$ 
at $\lambda=0$ to an interacting IR
fixed point with $c_{{\sst IR}}={\rm rank}\; H$ (for level $k=1$) at
$\lambda\sim 1$. Since the target bosonic string worldsheet we
find is critical only at the IR fixed point (for nonabelian $H$),
we conclude that for the conjecture articulated above to be valid,
the N=2 string coupling $\lambda$ should be fixed (presumably by
nonperturbative N=2 string considerations) to its IR value.
One can argue that $\lambda$ is quantized, by noting
that $1/\lambda^2$ multiplies the Wess Zumino term in target
space\foot{Actually, this point is somewhat confusing.  The 2+2d
theory has a Wess-Zumino term which is quantized via the Picard lattice
of the four-manifold; but it also induces the 1+1d Wess-Zumino
term via null reduction, whose quantization is more restricted.
We thank G. Moore for discussions of this issue.}. 
It would be interesting to understand what fixes it uniquely. 

\medskip

\noindent{{\sl Example 2}: N=(2,0) strings on the monster module.}

\smallskip

A well known example of a meromorphic c=24 CFT is the 
``monster module'' 
\ref\flm{I. Frenkel, J. Lepowsky and J. Meurman, 
Proc. Natl. Acad. Sci., USA, {\bf 81} (1984) 32566.},
which has no dimension one operators. When used to
construct a vacuum of the N=(2,0) string, it has {\it no} 
physical states (before compactification). The target
space dynamics is compatible (albeit in a rather trivial way)
with an N=(2,2) string worldsheet
in physical gauge, which also has no
transverse excitations.  

\medskip

\noindent{{\sl Example 3}: Type IIB target space strings
from the N=(2,1) string on the $E_8$ torus.}     

\smallskip

The vacuum described by \ztwo\ has eight bosonic and
eight fermionic physical states \vb, \vf, and N=(8,8)
global SUSY on the 2d target space.
The matter content and symmetry structure is that of the
type IIB string in static gauge \km. We'll verify later
that the dynamics agrees as well.

It is also worth mentioning that the type IIB construction
belongs to a larger class of critical 10d superstring 
target worldsheets, which can be obtained starting from the
underlying N=(2,1) string. In particular, in \kmo\ it has been 
shown how to get heterotic SO(32) and type I' strings by 
orientifolding.

\medskip

\noindent{{\sl Example 4}: Fermionic strings on group manifolds
from the N=(2,1) string.}     

\smallskip

The vacuum described by \zone-\gff\ provides an interesting
example of the conjecture discussed above. In \km\ it has 
been argued that this system describes in target space a bosonic
string worldsheet. However, this appeared problematic, since the
twenty-four 
scalars one finds transform in the adjoint of the group $H$, and it
is natural, as in example 1, to expect the dynamics to be described by
a WZW model for the group $H$ at some level k. For any finite k, the central
charge $c<24$, in apparent contradiction with the fact that we 
expect critical strings to emerge. 

The resolution of this contradiction could be the following:
as in example 1, it is natural to conjecture that the level
of the target space affine Lie algebra $\hat H$ is equal to
that of the worldsheet one; in this case it is $k=Q_H$, the
quadratic Casimir of $H$ in the adjoint representation.
The target space central charge is thus $c=D_H/2=12$, which indeed
seems non-critical. 

However, precisely for that level, the WZW model is equivalent
to a collection of (twenty-four) free fermions, $\{\theta^A\}$
in the adjoint representation of $H$, and furthermore, that collection
of fermions is invariant under a non-linear SUSY transformation
\eqn\ssu{\delta\theta^A=\epsilon f^{ABC}\theta^B\theta^C.}
This symmetry is not visible in the worldsheet construction of
\km\ since it acts in a highly non-linear fashion on the 
vertex operators of the twenty-four scalars in target space
\vvv.

We conclude that it is possible that \zone\ describes a critical 
ten dimensional fermionic string, rather than a non-critical bosonic 
one, in agreement with the conjecture.
Note that as in Example 1, there is close similarity 
between the worldsheet and target space structures.
In particular, an analog of \gff\ for the $\theta^A$ generates the target
SUSY transformations \ssu. The N=(2,1) string coupling must be
fixed to the value for which the WZW model is at its IR fixed point
in order that the target fermionic string is critical.
 
It is natural to expect the sum over spin structures of the
target space fermions $\theta^A$ to be the same as that for
the worldsheet fermions of the underlying N=2 heterotic
string, \zone, separately for left and right movers in target
space. Thus, this vacuum describes a non-supersymmetric
space-time 
compactification of the ten dimensional fermionic string to two
dimensions.
\medskip

\noindent{{\sl Example 5}: N=(2,2), (2,1) strings 
from the N=(2,1) string.}     

\smallskip

It is amusing to note that the N=(2,1) string is
``self-reproducing'', sometimes giving target space spectra which
are consistent with the N=(2,1) string itself or
other N=2 strings.  To get N=(2,2)
strings in target space, we split the twenty-four fermions
$\psi^A$ \gff\ into three groups of eight, and sum over
the spin structures of the three groups independently:
\eqn\zthree{Z(\tau)={1\over8}
\left[
\left({\theta_3\over\eta}\right)^4-
\left({\theta_4\over\eta}\right)^4-
\left({\theta_2\over\eta}\right)^4
\right]^3.}
Since the coefficient of $q^0$ in \zthree\ vanishes
sector by sector, it is clear that this model has
no physical states before compactification, just like the
N=(2,0) string vacuum in Example 2. As there, we 
interpret the dynamics as describing the worldsheet of an N=(2,2)
string in physical gauge\foot{Off shell, or after compactification,
the theory described by \zthree\ and that of Example 2 are clearly
inequivalent (\eg\ the former is supersymmetric in target space
while the latter is not), and possess non-trivial dynamics.
It would be interesting to understand them better.}.

To get the N=(2,1) string in target space, we start with the 
supersymmetric vacuum of Example 3, and orbifold by the $Z_2$ 
symmetry\foot{An alternative fermionic 
formulation of the same model
follows more closely the discussion of the N=(2,2)
target string above.  Split the twenty-four fermions
$\psi^A$ into three groups of eight, and sum over boundary
conditions such that the Ramond sector in one of the groups
is correlated with NS boundary conditions in the other two.
This makes it easier to evaluate the partition sum of the model,
given in the text.}
\eqn\zt{(x^a, \psi^a)\to -(x^a, \psi^a);\;\;\;a=4, \cdots, 11.}
The partition sum of the model is:
\eqn\zfour{Z(\tau)=-{3\over2}
\left({\theta_2\theta_3\theta_4\over\eta^3}\right)^4=-24.}
The spectrum includes the Ramond fields
\eqn\vw{\eqalign{
V_\alpha=&~\Sigma S_\alpha e^{ik\cdot x};\;\;\alpha=1, \cdots 8\cr
W_i=&~\Sigma\sigma_i e^{ik\cdot x};\;\;i=1, \cdots 16\cr
}}
where $S_\alpha$ are spin fields for $\psi^a$ (and transform
in the 8 of Spin(8)); $\sigma_i$ are the chiral twist fields
for the $\{x^a\}$, and $\Sigma$ is the dimension 1/2 spin
field for ghosts and longitudinal fermions $(\psi^0, \cdots, \psi^3)$.
The target space Dirac equation imposes the condition $k^+=0$.
Thus the spectrum of the model consists of
twenty-four free left moving
fermions \vw\ on the two dimensional 
target space, precisely the right
spectrum for interpreting the target space as an N=(2,1)
string worldsheet (again, in physical gauge). 
One expects a nonlinearly realized SUSY to act on these
fermions, but it clearly  shouldn't be visible via a linear action 
on the vertex operators \vw.  Finally, note that these vacua
(as well as the monster module, example 2) have the property that
boundary conditions on the fields require that
the null reduction always lie within the 2+2 longitudinal
dimensions; one has only target space strings and not membranes.

Hence, the 1+1d vacua of N=2 heterotic 
strings describe critical string worldsheets in their target
space, and their quantization is presumably equivalent to
the standard string one. As mentioned in the introduction,
the N=2 heterotic string construction appears to yield particularly
symmetric points in the moduli space of target strings, and 
it is not clear how to construct the full moduli space of
vacua in this framework. Consider, as an example, the 
construction of Example 1. It describes a critical
bosonic string compactified to 1+1 space-time dimensions.
It is well known that the moduli space of such theories
is $\MM\simeq SO(24,24)/SO(24)\times SO(24)$ 
(modulo the discrete
T duality group), while we are finding a unique theory
with no moduli, corresponding, presumably, to a 
separate compactification of the left and right movers
on an even self-dual twenty-four dimensional Niemeier 
lattice\foot{Of course, in 2d compactifications
of string theory one can not think of the Narain moduli
as fixed; their dynamics is described by a sigma model
with target space $\MM$. One can still describe states
in which the wave function of the moduli is centered
at some point in the classical moduli space.}.

In fact, this example is typical. Apparently, N=2 heterotic strings
lead to target string compactifications to 1+1 dimensions for
which left and right movers on the target worldsheet are {\it decoupled}.
Thus, example 3 is consistent with a superstring compactified 
on the $E_8$ torus, and in example 4 we find a 1+1 dimensional
fermionic string whose compact SCFT describes twenty-four
left and right
moving fermions whose spin structures are identified,
and summed separately for left and right movers. The target partition
sums in all the above cases appear to be the absolute values
squared of the worldsheet partition sums (A.3), \ztwo, \zone.
This is another example of the close connection of N=2 worldsheet 
and target physics mentioned above.

If this interpretation is correct\foot{As discussed in \refs{
\km, \kmo}, questions regarding sums over spin structures
and the shape of the compact manifolds in target space are
non-perturbative in the N=2 string coupling and thus are difficult
to address directly at present.}, 
perhaps the absence of Narain moduli of the target strings
is due to a symmetry that N=2 heterotic strings possess which
is only unbroken in the highly symmetric vacua described
above. It would be interesting to understand the
symmetry and find a more powerful
formalism that is capable of describing the full moduli
space of vacua.

What about the 2+1d vacua?
There, it is not known how to quantize theories of gravity 
coupled to matter, which makes it interesting to understand
the target dynamics of N=2 heterotic strings. 
In particular, choosing the 2+1 dimensional reduction of the
model in Example 3 above, yields a supersymmetric eleven
dimensional theory -- M-theory (type IIA string theory at
finite string coupling). By orbifolding this theory by a
symmetry that acts as the nontrivial outer automorphism 
of the N=2 algebra, one can construct \kmo\ vacua of 
M-theory compactified on $S^1/Z_2$ and thus study $E_8\times
E_8$ heterotic strings at finite coupling, as well as 
heterotic SO(32) and type $I^\prime$  strings.

\subsec{\it The target space geometry.}

Since all vacua of N=2 heterotic strings arise from 2+2 dimensional
theories by different null reductions, it is natural to study 
the target space dynamics by first relaxing the constraint \phst,
understanding the resulting 2+2 dimensional dynamics, and
then restoring the null reduction.

The first question one needs to address is what is the 
geometrical meaning of the scalars \vbos, \vb, \vvv\ and
gauge fields \gf, \gfone\ in 2+2d.  The answer to that question
is known 
\nref\hullwit{C. Hull and E. Witten, \pl{160}{1985}{398}.}%
\nref\hull{C.M. Hull, \pl{178}{1986}{357};%
\np{267}{1986}{266}.}%
\refs{\hullwit,\hull} (see also \ovtwo): 
Performing the $\theta, \bar\theta$ integrals
one finds that the scalars $V_\phi^a$ can be thought of as parametrizing
a self-dual gauge field:
\eqn\aj{
\eqalign{
A_j=&~i\partial_j\Omega^{-1}\Omega\cr
A_{\bar j}=&~i\partial_{\bar j}\Omega \Omega^{-1}\cr
}}
where $\Omega=\exp\left(\phi^a t^a\right)$, and 
$t^a$ are Hermitian generators of $H$. 
In the abelian case we have:
\eqn\ajtwo{
\eqalign{
A_j^a=&-i\partial_j\phi^a\cr
A_{\bar j}^a=&i\partial_{\bar j}\phi^a\quad \cr
}}
The gauge field $V_K$ \gfone\ 
parametrizes a Hermitian metric with torsion. 
Its contribution to the N=(2,1) string worldsheet
Lagrangian is described by the deformation
\eqn\grav{\int d^2z\int d\theta\int d^2\bar\theta \ K_\mu(x)DX^\mu.}
Defining
\eqn\fmunu{F_{\mu\nu}=\partial_\mu K_\nu-\partial_\nu K_\mu}
and performing the $\theta, \bar\theta$ integrals, one finds that
$F_{\mu\nu}$ parametrizes a metric with torsion:
\eqn\metr{g_{\mu\nu}=\half\{I, F\}_{\mu\nu};\;\;
B_{\mu\nu}=\half[I, F]_{\mu\nu}
.}
In complex coordinates, $I_{i\bar j}=\delta_{i\bar j}$
we have the metric and $B$ field:
\eqn\gb{
\eqalign{
g_{i\bar j}=&~\partial_i K_{\bar j}-\partial_{\bar j} K_i\cr
B_{i\bar j}=&~\partial_i K_{\bar j}+\partial_{\bar j} K_i\cr
}}
We will typically expand around a flat background,
$g_{\ij}=\eta_\ij + F_{i\bar j}$.
We see that the CFT on the target worldsheet describing different
kinds of strings is generalized in an interesting way in passing
to 2+2d, to a non-linear self-dual system of gauge fields \aj, and gravity
\gb. In some cases, this system should be supersymmetrized \supch.

It should be very interesting to understand the dynamics of this system
directly in 2+2d. Therefore we will examine in the
next section the 2+2d classical 
dynamics describing the target space fields of N=2 heterotic strings.
We conclude this section with some brief comments 
on compactification of N=2 heterotic strings. 

\subsec{Compactified N=2 heterotic strings.}

Consider an N=2 heterotic string compactified on a circle of radius
R, $\IR^{2,2}\to \IR^{2,1}\times \s^1$. One of the most striking features
of N=2 heterotic strings is the qualitative difference between the 
spectra of the finite and infinite R theories.
As explained in previous sections, for infinite R the spectrum 
includes a finite collection of field theoretic 
degrees of freedom (self-dual
gauge fields and metric). For any finite R, we find a rich string
theoretic spectrum with an exponential density of states\foot{
The role of winding N=2 heterotic strings has been also
discussed in 
\ref\barbon{J. Barbon and M. Vasquez-Mozo, hep-th/9605050.}.}.

Indeed, while the right movers are forced by the N=2 physical
state conditions to remain in their ground state, one can for finite
radius excite the whole tower of left moving oscillators, in mixed
momentum/winding sectors. The mass spectrum satisfies:
\eqn\mn{M^2=({n\over R}+{mR\over 2})^2=({n\over R}-{mR\over 2})^2
+2N}
where $n,m$ are the momentum and winding around the $\s^1$,
and $N$ the level of left-moving oscillator excitation\foot{We 
are suppressing momenta along the chiral, left-moving torus,
which do not lead to a significant modification of the following
discussion. We will return to them later in the paper.}. 
By \mn,
\eqn\nnm{N=nm.}
In the $R\to\infty$ limit all states with $m\not=0$ go to infinite
mass, and we are left with the field theoretic spectrum described
above. For any finite R, the density of states is exponential. In a 
sector with given $n,m$, \mn, \nnm\ imply that the mass $M$ is fixed,
$M\sim N$, and since the number of states at level $N$ 
is of order $N^\alpha\exp(\beta\sqrt N)$,
we have for given $n,m$ of the order of $M^\alpha\exp(\beta\sqrt M)$
states.
It is important for the dynamics -- in particular the
triviality of the S-matrix -- that this number of states,
while of course very large by field theoretic standards,
is smaller than that in ordinary string theory (where
the number of states with mass $M$ grows like
$\exp(\beta M)$).  

The {\it total} density of states {\it is} string theoretic. Indeed,
for states with a certain large mass $M\gg R, 1/R$, 
the density of states
is dominated by the sectors with $n/R\simeq mR/2\simeq M/2$, and
by \nnm, $N\simeq M^2/2$. The density of states grows with 
mass as $\rho(M)\simeq M^\alpha e^{\beta M}$.

Thus, N=2 heterotic strings have the remarkable property
that at infinite radius physics is field theoretic, while
for any finite radius the system exhibits a Hagedorn transition
and has a much richer dynamics. There are two other contexts
in recent discussions of string theory that exhibit similar features;
both are probably related to N=2 heterotic strings.

\item{1)} {\it The physics of BPS states in string theory.}
In ten dimensional uncompactified (perturbative)
superstring theory, the
only states which are killed by half of the supercharges
(and thus belong to short representations of space-time SUSY)
are massless. Compactification on one or more circles gives rise
to a rich spectrum of Dabholkar-Harvey states
\ref\dh{A. Dabholkar and J. Harvey, \prl{63}{1989}{478};
A. Dabholkar, G. Gibbons, J. Harvey and F. Ruiz Ruiz, 
\np{340}{1990}{33}.}\ with a spectrum analogous to \mn. As we'll
discuss below, this analogy is not accidental -- one can think
of N=2 heterotic strings as describing the sector of BPS states of type
II / (1,0) heterotic strings.

\item{2)} {\it String duality.} If one replaces the radius by the 
string coupling in the above considerations, one arrives at a picture
familiar from string duality. As the string coupling goes to zero, the
magnetic and dyonic strings go to infinite mass and decouple. For any 
finite coupling, a duality symmetric formulation of any string theory
would presumably include electric, magnetic and dyonic 
degrees of freedom. It is not
understood what those are in general (and it is one of the main
motivations of our work to gain a better understanding of them),
but a recent proposal for understanding S-duality in heterotic
string theory on $T^6$ \dvv\
provides a concrete realization in which magnetic and dyonic 
strings are realized as winding and mixed momentum-winding
modes of an underlying string, whose target space is (roughly)
the heterotic worldsheet. 

\noindent
To summarize, we propose that the peculiar behavior of N=2 heterotic
strings upon compactification is related to similar phenomena in the
context of BPS physics and string duality. We will return to both
later.

\newsec{The target space action}

The physical degrees of freedom of the metric, $B$-field, and
dilaton are encoded in the one-form potential $K$ 
\grav-\gb; those of
the gauge fields in a Yang scalar $\phi^a$ \aj, \ajtwo; and 
the Rarita-Schwinger fields are reduced to the derivative of
a set of spin one-half fields $\vartheta$.  In this section
we will derive all of the purely bosonic terms in the target
space action governing the dynamics of these fields, as well 
as a few of the terms involving fermions.  
We start by reviewing some known features of the dynamics
of N=2 strings. 

One of the remarkable features of N=2
strings is the fact that while they generically have
non-vanishing three-point couplings, higher-point functions
vanish on-shell. 
This phenomenon is due to a clash between the following
two properties: 
N=2 string amplitudes exhibit the usual string theoretic
exponential fall off at large transverse momentum transfer;
at the same time they have a finite number of field theoretic
degrees of freedom (before compactification). 
The incompatibility of the two forces all scattering
amplitudes that depend non-trivially on kinematic invariants
($N\ge 4$ point functions) to vanish. 

After compactification, one might have expected to find
non-trivial higher-point functions since the spectrum
is no longer field theoretic -- in fact, as we've seen
in section {\it 2.4} the total density of states is
string theoretic. However, the density of states that
is relevant for the behavior of amplitudes
at large momentum transfer 
is the density of states with {\it fixed momentum and
winding}. That density of states was shown in section
{\it 2.4} to grow like $\exp(\sqrt M)$ and hence, is
insufficient to give Regge behavior of amplitudes.
Thus, despite the string theoretic spectrum, compactified
N=2 strings also have vanishing $N\ge4$ point functions. 

The fact that most $N$ point functions vanish does not
in general mean that the target space action is cubic.
Iterating the three-point function to calculate an
on-shell S-matrix element generically gives a non-zero
{\it local} four-point coupling. To reproduce the 
vanishing four-point S-matrix of the string, one
then has to add an explicit local four-point interaction
to cancel the iterated three-point coupling. Repeating
this procedure, one generically finds a non-polynomial
action, which in principle can be determined recursively
from the three-point function as indicated above. In
practice, there are subtleties having to do both
with calculational difficulties at high orders, and
the ambiguity of inferring an off shell irreducible
vertex from the knowledge of an on-shell counterterm
(the two difficulties are of course related).
As we'll see, symmetries provide an important guide
and sometimes allow one to solve the problem.

The simplest case is that of the N=(2,2) string \ovone.
The iteration of the three-point function vanishes on-shell due to  
kinematic identities in 2+2 dimensions; there is no
local four-point contact term, and all higher-point
contact interactions vanish as well.  At the same time,
the N=(2,2) sigma model in four dimensions has global N=(4,4)
supersymmetry (one easily constructs the additional currents
by spectral flow \ref\eoty{T. Eguchi, H. Ooguri, A. Taormina,
and S.-K. Yang, \np{315}{1989}{193}.}), for which
the beta functions are exact at one loop.  All of this information
points to the cubic Plebanski action for self-dual gravity
as the target space theory \ovone:
\eqn\pleb{
  S_{\sst N=(2,2)}=\int I\wedge\d\phi\wedge\dbar\phi
	+\d\phi\wedge\dbar\phi\wedge\d\dbar\phi\ ,
}
where the \kahler\ form is $k=I+\d\dbar\phi$.
The variation of this action yields the equation $det[g]=1$;
the beta function equations follow from $R=\d\dbar log\det[g]=0$.

The situation for N=(2,1) and N=(2,0) strings is more complicated.
The analysis of Ooguri and Vafa \ovtwo\ indicates that, in order to
enforce the vanishing of the S-matrix, contact terms must be introduced
at each order to cancel the iteration of lower-point S-matrices.
This is in agreement with the fact that 
the sigma-model has only global N=(4,1) or N=(4,0) supersymmetry,
so the beta-functions might receive contributions to all orders in
perturbation theory (\cf\ \ref\fourzero{C. Callan, J. Harvey,
and A. Strominger, \np{359}{1991}{611}; {\it Proceedings, String theory
and Quantum Gravity}, Trieste 1991, hep-th/9112030.
P. Howe and G. Papadopoulos, hep-th/9203070, \np{381}{1992}{360}.}).
In this section we will determine these counterterms.
Our analysis will concentrate on N=(2,1) strings; our
methods are less powerful for N=(2,0) strings, and we will highlight
a few of the differences along the way.

The basic idea is to use the complementary information about
the target space action encoded in the S-matrix elements
and the sigma model beta functions;
both encode the target space action, but in different ways.
The $n$-point S-matrix yields terms with any number of derivatives
and $n$ fields, while the $m^{\rm th}$ order in the loop expansion
of the beta-functions gives all powers in fields with $m$ derivatives.
We will see that by comparing the two at low orders
in the expansion, and in particular by using the fact that
the full S-matrix vanishes, 
one can in some cases deduce the entire structure.
We start with a discussion of the gravitational sector.

\subsec{{\it The gravitational sector}}

The three-point S-matrix of the gravitational modes \grav,
with momenta $k_r$ and polarizations $\xi_r$, $r=1,2,3$,
is given by
\eqn\threept{
  \langle V_K(1) V_K(2) V_K(3) \rangle =
  [k_1\cdot I\cdot k_3]\times[
  (\xi_1\cdot\xi_2)(k_1\cdot\xi_3)+
  (\xi_1\cdot\xi_3)(k_3\cdot\xi_2)+
  (\xi_2\cdot\xi_3)(k_2\cdot\xi_1)]\ .
}
Equation \threept\ factorizes into a right-moving contribution
which is the standard one for N=2 strings, and a
left-moving one which is identical to that of a gauge field in
N=1 string theory.  Indeed, the left-moving part of \grav\
is just that of an N=1 gauge boson, with the standard gauge
invariance $K_\mu\rightarrow K_\mu+\d_\mu\eps$; here this
symmetry corresponds to a redundancy of the description of the
metric in terms of $K_\mu$ \metr\foot{In \ovtwo, a nonabelian
form of this symmetry is suggested; this transformation is not
compatible with the vanishing of the S-matrix.  One can also see
that the symmetry is abelian by an examination of the sigma-model
action.}.
The form of the three-point interaction that gives \threept\
on-shell and preserves the above gauge invariance is:
\eqn\lthree{
  \LL_3=\hf F_{\mu\nu}F_{\nu\lam}F_{\lam\rho}I_{\rho\mu}
	-\coeff18 F_{\alpha\beta}F_{\beta\alpha}
		I_{\mu\nu}F_{\nu\mu}\ ,
}
where $F_{\mu\nu}$ determines the geometry via \fmunu-\gb.
A useful observation for the subsequent discussion is that,
in coordinates that diagonalize $I_{\mu\nu}$,
$\LL_3$ can be rewritten as
\eqn\lmod{
  \LL_3=F_{i\bar j}F_{j\bar k}F_{k\bar i}
	-\hf F_{i\ibar}F_{j\bar k}F_{k\bar j}\ .
}
In particular, it is independent of $F_{ij}$,
$F_{\bar i\bar j}$ when $I$ is diagonal.  For simplicity,
let us temporarily work with such a complex structure.

\nref\oneloop{G. Bonneau and G. Valent, \cqg {11}{1994}{1133}; 
hep-th/9401003.}%
Another important fact about \threept, \lthree\ is that the
cubic vertex has the qualitative form $\LL_3\sim \xi^3 k^3$,
\ie\ it is cubic in fields and momenta\foot{In contrast,
the analogous three-point function in the N=(2,0) string has a piece
proportional to $\xi^3 k^5$, which would invalidate the simplest
version of the subsequent discussion.  It is not difficult to see
that the origin of this term is the gravitational Chern-Simons
contribution to the sigma-model anomaly, which is
cancelled in the (2,1) string by the left-moving superpartners 
$\psi^\mu$ of the target space coordinates
$x^\mu$; on the other hand,
this anomaly does contribute
to the (2,0) theory by the usual shift of the $B$-field.}.
Power counting then implies that iterating the cubic
vertex leads to a four-point coupling that behaves as
$\xi^4 k^4$ (two factors of $k^3$ from the vertices and
$1/k^2$ from the propagator), and more generally the 
$n$-point coupling will go like $\xi^n k^n$.  As mentioned above,
the $3<n$-point S-matrix elements all vanish;
hence the above $n$-point couplings must be 
cancelled by explicit, local higher order
contributions to the effective action.  The power-counting
argument implies that the necessary higher-order couplings
go like $\LL_n\sim (F_{\mu\nu})^n$ and hence {\it are given exactly
by a one-loop calculation} in the N=(2,1) sigma model \grav\
(recall that the sigma model metric is essentially
$g_{\mu\nu}=\eta_{\mu\nu}+F_{\mu\nu}$).
Fortunately, the necessary calculations have already been
done \refs{\hull,\oneloop}; let us recall the results.  
The connection with torsion $\Gamma^\mu_{\nu\lam}$
for \gb\ takes the form
\eqn\connec{\eqalign{
  \Gamma^i_{jk}= &~ g^{i\bar\ell}g_{k\bar\ell,j} \cr
  \Gamma^i_{j\bar k}= &~ g^{i\bar\ell}[g_{j\bar\ell,\bar k}
	-g_{j\bar k,\bar \ell}]\ .
}}
The standard beta-function equations 
$R_{\mu\nu}[\Gamma]=\nabla_\mu\nabla_\nu\Phi$
can be integrated once
or twice, and yield the set of equations
\eqn\inteq{\eqalign{
  \Gamma_\mu=&~ 0\cr
  \log\;\det[g_{i\bar j}]=&~2\Phi\ ,
}}
where $\Gamma_\mu=\Gamma^{\rho}_{\nu\mu}I^\nu_\rho$, and
$\Phi$ is the dilaton.  The second equation in \inteq\
fixes the dilaton in terms of the metric.  The first equation
serves as an equation of motion for \gb.  One can show that,
treating $K_\mu$ \grav\ as independent variables, this equation
follows from the action\foot{This action for the gravitational
dynamics was independently arrived at by Chris Hull
\ref\hullpriv{C.M. Hull, private communication.}.}
\eqn\gract{
  \LL_g=T\sqrt{\det[\eta_{i\bar j}+2F_{i\bar j}]}\ .
}
The expression for a general complex structure is easily deduced
by an O(2,2) rotation.
The tension $T$ is related to the
N=2 string coupling, $T=1/g_{{\rm str}}^2$.
Note that the determinant here is of a $2\times2$ matrix, not
a $4\times4$ matrix.
Another form of the action \gract\ incorporates the dilaton
equation (the second of equations \inteq), and will be useful
in our discussion of the fermionic terms below:
\eqn\bltone{
  \LL_g=e^{-\Phi}{\det[g_\ij]}+T^2e^{\Phi}\ .
}
Eliminating $\Phi$ by its algebraic equation of motion gives
back \gract.
Expanding $\LL_g$ (equation \gract) in $F$, setting
$T=1$, we find for the first few
nontrivial orders (and up to total derivatives)
\eqn\explg{
  \LL_g= -\coeff14 F_{\mu\nu}F_{\mu\nu} -
	\coeff14 F_{\mu\nu}\widetilde F_{\mu\nu}
	+f A_{\ij}A_{j\ibar} -\hf(A_\ij A_{j\ibar})^2
	-f^2 A_\ij A_{j\ibar} +\ldots\ ,
}
where we have introduced the notation
\eqn\deffA{\eqalign{
  f=&~F_{i\ibar}\cr
  A_\ij=&~F_\ij -\hf \eta_\ij f\ .
}}
and $\widetilde F^{\mu\nu}={1\over2}\epsilon^{\mu\nu\lambda
\rho}F_{\lambda\rho}$.
The kinetic term in \explg\ gives the correct propagator
for $K_\mu$, while the cubic interaction $fA^2$
can be checked to be equivalent to \lmod.  We will check 
later that the quartic terms in \explg\
are exactly what is needed to cancel the effect of iterating
the three-point coupling.

Note that \gract\ can be written as $\LL_g=[\det g_{\mu\nu}]^{1/4}$,
where $g$ is the $4\times4$ Hermitian metric.  It is
curious that $\LL_g$ is not a density; the reason is that
one has fixed complex coordinates, and furthermore one must make
a holomorphic coordinate transformation to remove integration
`constants' of the form $h(x)+\bar h(\bar x)$ in \inteq.
The residual symmetry group of \gract\ is thus holomorphic
area-preserving diffeomorphisms
\eqn\symgr{
  x^i\rightarrow f^i(x^j)\quad ;\qquad |\det \d_j f^i|=1\ .
}
It is interesting to note the progression of target space theories
as the worldsheet supersymmetry is increased:
\eqn\comparison{\eqalign{
  R_{\mu\nu}=\nabla_\mu\nabla_\nu\Phi+\ldots\ ,\qquad&N=(1,1)\cr
	\Gamma_\mu=\Gamma^{\rho}_{\nu\mu}I^\nu_\rho=0\qquad ,
		\qquad&N=(2,1)\cr
	\det[g]=1\qquad ,\qquad\qquad\qquad&N=(2,2)\ .\cr
}}

\subsec{{\it Adding the self-dual gauge fields}}

The vertex operators probing the `internal space' of 
N=2 heterotic strings describe self-dual gauge fields
(\cf\ section {\it 2.3}) \ovtwo.
The background field with canonical kinetic energy in the target
space dynamics is the Yang scalar $\phi^a$ of eq. \aj, which
parametrizes 
self-dual configurations.
The linearized self-duality
equations then reduce to $\d_i\d_\ibar\phi^a=O(\phi^2)$.
The first nonlinear term on the RHS arises from the cubic
S-matrix element
\eqn\threegauge{
  \langle V_\phi(1) V_\phi(2) V_\phi(3) \rangle =
	[k_2\cdot I\cdot k_3]\times[\zeta_1^a\zeta_2^b\zeta_3^c f_{abc}]\ , 
}
corresponding to the interaction Lagrangian
\eqn\lthreegauge{
  \LL_3^{{\sst gauge}}=\phi^a\d_\mu\phi^b\d_\nu\phi^c I^{\mu\nu}f_{abc}\ .
}
The gauge fields also couple to the gravitational sector;
the leading interaction is
\eqn\gagagr{
  \langle V_\phi(1) V_\phi(2) V_K(3) \rangle =
	[k_2\cdot I\cdot k_3]\times[\zeta_1\cdot\zeta_2\ \xi_3\cdot k_1]\ ,
}
arising from the interaction Lagrangian
\eqn\lthreegaugegrav{
  \LL_3^{{\sst gauge-grav}}=
	(F_{\mu\lam}I_{\lam\nu}-\coeff14 \eta_{\mu\nu}
	F_{\alpha\beta}I_{\beta\alpha})\d_\nu\phi^a\d_\mu\phi^a\ .
}
Note that this last expression can be
obtained from \lthree\ by extending all indices to twelve dimensions,
while keeping the coordinates four-dimensional (and of course
$I_{\mu\nu}$ nonzero only in four dimensions).

\nref\dns{S. Donaldson, Proc. Lond. Math. Soc., {\bf 50} (1985) 1;
V. Nair and J. Schiff, \pl{246}{1990}{423};
\np{371}{1992}{329}.}%
\nref\lmns{A. Losev, G. Moore, N. Nekrasov, and S. Shatashvili, 
hep-th/9509151; hep-th/9606082.}%
Ooguri and Vafa \ovtwo\ analyzed the dynamics of the gauge 
field, ignoring gravity.  This is self-consistent since 
the gauge-gravitational interactions are higher-order in derivatives.
The vanishing
of the four-point S-matrix (see also 
\ref\parkes{A. Parkes, \np{376}{1992}{279}})
and the self-duality equations (which are equivalent to the
one-loop gauge field beta-function equations in four dimensions)
lead to the effective action
\eqn\sfourwz{\eqalign{
  S_{{\sst gauge}}=&~\int I\wedge \omega_{{\sst WZ}}\cr
  \omega_{{\sst WZ}}
	=&~\int_0^1 dt\;{\rm Tr}
		[\dbar\phi \wedge e^{-t\phi}\d e^{t\phi}]\cr
	=&~{\rm Tr}\biggl[\hf\dbar\phi\wedge\d\phi+\sum_{n=3}^\infty
	\frac{1}{n!}\dbar\phi\wedge
		[\cdots[[\d\phi,\phi],\phi]\cdots]\biggr]\ ,
}}
which is the four-dimensional Wess-Zumino action 
\refs{\dns, \lmns}.
To discern the coupling of the gauge sector to gravity,
it is simpler to consider the abelian case \ajtwo.
A power-counting argument similar to that employed in
the previous subsection implies that $n$-point functions
must depend on $\phi^a$ only via $(\d_\mu\phi^a)^n$, \ie\
$(A_\mu^a)^n$.  Thus, to determine the contribution of
the $\{\phi^a\}$ to the target space action, it is
sufficient to consider the case of constant (or very slowly
varying) gauge fields $A_\mu^a$.  In that limit, the only
effect of the gauge fields is a shift (due to the sigma
model anomaly) 
\nref\hulltown{C.M. Hull and P.K. Townsend, \pl{178}{1986}{187}.}%
\refs{\hullwit,\hulltown} 
\eqn\shift{\eqalign{
  g_\ij&\rightarrow g_\ij+A_i^a A_{\bar j}^a\cr
  H=dB&\rightarrow H+\omega_3^{{\sst YM}} \ .
}}
Thus the action for all propagating bosonic fields 
in this theory is
\eqn\lfull{\eqalign{
  \LL=&~T\sqrt{\det[g_\ij]}\cr
  g_\ij=&~\eta_\ij+2F_\ij+2\d_i\phi^a\d_{\bar j}\phi^a\ .
}}
This is our main result: The action of 2+2 M-branes
bears an intriguing resemblance to the 
Nambu-Goto/Dirac-Born-Infeld actions describing D-branes.
Specifically, it looks like a complexification of the effective
action describing a D-string, but with two differences:
1) \lfull\ is the {\it exact} target space action of the
N=(2,1) string, while its analog for D-strings receives
corrections both from higher orders in $\alpha'$, 
and from interactions
with closed string states that live in the bulk of spacetime;
2) the interpretation of the fields is different -- the gauge
field on the worldsheet of a D-string is replaced here by
a Hermitian metric potential $K_\mu$, while the scalars $\phi^a$
parametrizing transverse motion of the D-string here
arise as self-dual $U(1)^8$ gauge fields \ajtwo.

Expanding \lfull\ to quartic order in the fields, we find
($\LL_g$ is given in equation \explg): 
\eqn\lll{\eqalign{
  \LL_{{\rm full}}&~= \LL_g+\hf(\d_\mu\phi^a\d_\mu\phi^a)
	-2A_\ij\d_j\phi^a\d_\ibar\phi^a\cr
	&~ +\hf(\d_i\phi^a\d_\ibar\phi^a)^2
	-\d_i\phi^a\d_{\bar j}\phi^a\d_j\phi^b\d_\ibar\phi^b
	+2fA_\ij\d_j\phi^a\d_\ibar\phi^a
	+A_\ij A_{j\ibar}\d_\ell\phi^a\d_{\bar\ell}\phi^a\ .
}}
The cubic interaction in \lll\ is the same as that
obtained from a direct examination of the S-matrix 
\lmod, \lthreegaugegrav.
The quartic terms will be verified below as well.

The generalization of \lfull\ to the nonabelian case is 
straightforward; one simply replaces $g_{\ij}$ by
\eqn\nonab{
  g_\ij=\eta_\ij+2F_\ij+4\omega_\ij^{{\sst WZ}}\ ,
}
where $\omega_\ij^{{\sst WZ}}$ are the components of the Wess-Zumino
two-form of \sfourwz.  The lagrangian $\LL=T\sqrt{|g_\ij|}$
then reproduces the two-derivative term \sfourwz\ deduced
by \ovtwo, as well as the correct cubic S-matrix and
four-point contact terms. It is important to emphasize that
the simple analysis performed above has led us to a rather
striking prediction -- that the lagrangian $\LL$ has in fact
a vanishing S-matrix to all orders in the fields! As we'll 
see momentarily, checking this statement even to quartic order
is rather non-trivial, and we haven't been able to prove
this claim to all orders by a direct evaluation of
Feynman diagrams.  It is important to understand the reason
for the vanishing; perhaps the symmetries \symgr\ can be
utilized to explain this remarkable result (\cf\ 
\nref\popov{A.D. Popov, M. Bordemann, H. Romer,
hep-th/9606077; \pl{385}{1996}{63}.}\refs{\ovone,\popov}
for a discussion of the N=(2,2) case). 

The canonical dimensional reduction of the action \nonab\
to two dimensions obtained by setting
the imaginary part of $x_i$ to zero,
gives the Nambu-Goto action for a string whose
transverse coordinates lie in a group manifold.  In the discussion
of section 2 we saw that the Wess-Zumino term, determining the
level of the resulting affine Lie algebra, played an important role. 
Can there be such a two-dimensional Wess-Zumino term 
in the target space theory?  
One can see such a term directly from the 
four-dimensional Wess-Zumino
action of \sfourwz, using the general complex structure \iij.
The gauge vertex three-point function \lthreegauge\
in a theory with a non-trivial complex structure \iij\ (i.e.
with $b\not=0$) has a term:
\eqn\twodwz{
  b\;\phi^a\d_i\phi^b\d_j\phi^c \eps_{\ibar\bar j} f_{abc}
	+{\rm h.c.}\ ;
}
with the canonical null reduction $x^i=x^\ibar$, this is precisely
the leading term in the expansion of the two-dimensional Wess-Zumino
term in powers of $\phi$.

\subsec{{\it A check of the action}}

We will now perform a check of the action \lfull\
to quartic order.  The cubic interactions in the Lagrangian are
(see \deffA\ for the notation)
\eqn\lcubic{
  \LL^{(3)}=f A_\ij A_{j\ibar} - 2A_\ij \d_j\phi^a\d_\ibar\phi^a\ .
}
Iterating the three-point coupling yields four-point couplings
of the form $(K_\mu)^4$, $(K_\mu)^2(\phi^a)^2$, and
$(\phi^a)^4$.  If the Lagrangian \lfull\ is correct, 
two things must happen.  First, all poles in the S-matrix must cancel
on-shell.  Second, any remaining local terms 
in the iterated three-point function must cancel against the
explicit four-point couplings in \explg, \lll,
such that the total four-point S-matrix vanishes.

To perform the check, one uses the propagators
\eqn\propg{\eqalign{
  \langle K_\mu(k) K_\nu(-k)\rangle=&~ \frac{\eta_{\mu\nu}}{k^2}\cr
  \langle \phi^a(k) \phi^b(-k)\rangle=&~ \frac{\delta^{ab}}{k^2}\ .
}}
Using these propagators, one can show that for any two traceless
$2\times2$ matrices $M_{\ij}$, $N_\ij$,
\eqn\contrc{
  M_\ij\langle A_{j\ibar} A_{\ell\bar m}\rangle N_{m\bar\ell}
	=\hf M_\ij N_{j\ibar}\ .
}
The poles in the propagator \propg\ cancel and one gets a
local expression \contrc.  Using the identity \contrc,
one can show that all the terms in $\langle\LL^{(3)}\LL^{(3)}\rangle$
which involve contractions of $A_\ij$ with itself add up to
\eqn\reslt{
  \langle\LL^{(3)}\LL^{(3)}\rangle_{{\sst \langle AA\rangle \rm terms}}
	=-2\LL^{(4)}\ ,
}
where 
\eqn\ress{
\eqalign{
\LL^{(4)}&=-\hf(A_\ij A_{j\ibar})^2
        -f^2 A_\ij A_{j\ibar}\cr
&+\hf(\d_i\phi^a\d_\ibar\phi^a)^2
        -\d_i\phi^a\d_{\bar j}\phi^a\d_j\phi^b\d_\ibar\phi^b
        +2fA_\ij\d_j\phi^a\d_\ibar\phi^a
        +A_\ij A_{j\ibar}\d_\ell\phi^a\d_{\bar\ell}\phi^a\cr}} 
is the quartic interaction in \explg, \lll.
This leaves terms obtained by contracting $\langle f A_\ij\rangle$
or $\langle\phi^a\phi^b\rangle$.  
These all turn out to cancel amongst themselves.
For example, consider the terms
\eqn\ggss{
  2 A_\ij A_{j\ibar}\frac1\lform
	\d_{\kbar}\d_\ell\phi^a\d_k\d_{\lbar}\phi^a +
  A_\ij\d_\ibar\d_j\phi^a\frac1\lform A_{k\lbar}\d_\kbar\d_\ell\phi^a
}
that appear in the scattering of two gravitons and two scalars.
The first term comes from a contraction $\vev{f A_{k\lbar}}$,
the second from $\vev{\phi\phi}$.
One can show that the combination of these
terms vanishes on-shell, by repeated use of the kinematic 
identities \refs{\ovone,\parkes}
\eqn\ident{
  \frac{(k_1\cdot\kbar_2)\;(k_4\cdot\kbar_3)}
	{s_{12}}
  =-\frac{(k_1\cdot\kbar_3)\;(k_4\cdot\kbar_2)}
	{s_{13}}
  =-\frac{(k_1\cdot\kbar_4)\;(k_2\cdot\kbar_3)}
	{s_{14}}-k_1\cdot\kbar_3\ ,
}
and other identities
obtained by antisymmetrizing on three holomorphic or three
antiholomorphic indices.  
Here $s_{\alpha\beta}=k_\alpha\cdot {\bar k}_\beta + 
k_\beta\cdot {\bar k}_\alpha$ are the Mandelstam variables.
This sort of highly nontrivial cancellation is one of the
seemingly miraculous properties of self-dual gauge systems
in four dimensions.
Given this result, equation \reslt\ then implies that
the quartic terms obtained by iterating the three-point function
cancel the explicit four-point couplings in \lfull.

\subsec{{\it Fermionic terms}}

So far we have concentrated on the bosonic terms in the action. 
The interaction arising from the cubic S-matrix 
involving fermions\foot{The origin of the fermions is a four-dimensional
Rarita-Schwinger field $\chi_\mu^{\;\alpha}=I_\mu^\nu\d_\nu\ferm^\alpha$.
The fermion $\ferm$ is a potential for it, just as $K_\mu$, $\phi^a$
are potentials for $g$, $b$, and $A$.}
\eqn\fermcub{
  \vev{V_\ferm(1)V_K(2)V_\ferm(3) + V_\ferm(1)V_\phi(2)V_\ferm(3)}= 
	[k_2\cdot I\cdot k_3]\times [\bar u_1(\xi_2^\mu\Gamma_\mu
	+\zeta_2^a\Gamma_a)u_3]
}
can be obtained from the cubic Lagrangian:
\eqn\ltf{
\LL^{(3)}=\bar\ferm\Gamma^\mu\partial_\nu\ferm I^{\nu\lambda}
F_{\lambda\mu}+\bar\ferm\Gamma^a\partial_\nu\ferm I^{\nu\lambda}
\partial_\lambda\phi^a\ .
}
By analogy with D-strings and D-twobranes,
we would expect the fermions to lead to two modifications:

\item{1)} In \lfull\ we should replace:
\eqn\modf{\eqalign{
  \partial_i\phi^a\rightarrow&~\Pi_i^a\equiv\partial_i\phi^a-
	\bar\ferm\Gamma^{a}\partial_i\ferm\cr
  \partial_\ibar\phi^a\rightarrow&~\Pi_\ibar^a\equiv\partial_\ibar\phi^a-
	\bar\ferm\Gamma^{a}\partial_\ibar\ferm\cr
}}

\item{2)} The volume term \lfull\ should be 
corrected by a Wess-Zumino term, which will be needed
for the $\kappa$-symmetry expected in a covariant formulation.

\noindent
The cubic interaction \ltf\ is compatible with the extension \modf;
making this substitution in \lfull\ (or its expansion \lll), 
one finds complete agreement to this order for the interaction 
of the fermions with the transverse scalars $\phi^a$.  
The first term in \ltf\ also appears in the expansion of \lll\
about a general complex structure, with the assumption that the 
$\eta_{i\jbar}$ term in \lfull\ arises as the expectation
value of a covariant $\Pi_\mu^a$ (\ie\ including longitudinal
scalars $\phi^\mu$) in static gauge, $\vev{\Pi_\mu^a}=\delta_\mu^a$.
However, we have not checked the precise coefficients.

On very general grounds, one would expect the target theory
to have a local fermionic invariance.  
String theory always contains gravity in its target space;
supersymmetric string theories always have a local fermionic symmetry
whose generators close on the diffeomorphism generators.  Usually
this local fermionic symmetry is gauged supersymmetry; in the
present case, the natural candidate is $\kappa$-symmetry\foot{String
theory presents these symmetries in a fixed gauge.  One is used
to conformal gauge on the worldsheet admitting a linearized
gauge symmetry in target space; in the N=2 string, the additional
BRST constraints on the gauge parameter eliminate such a transformation.}.
A concrete indication that the action results from the gauge-fixing
of a $\kappa$-symmetry comes from the linear spacetime supersymmetry
of the N=(2,1) string.  A superspace completion of the form
\modf\ results in a theory with global fermionic invariance
\eqn\stsusy{\eqalign{
  \delta\ferm=&~\eps\cr
  \delta\phi^a=&~\bar\eps\Gamma^a\ferm\ ;\cr
}}
in static gauge, this symmetry can be used to cancel the 
inhomogeneous part of the global remnant of the
$\kappa$-symmetry transformation
\eqn\kappasym{\eqalign{
  \delta\ferm=&~(1+\Gamma_{p+1})\kappa\cr
  \delta\phi^a=&~(\delta\fermbar)\Gamma^a\ferm\ ,\cr
}}
leading to a linear supersymmetry on the worldvolume.  
Here $\Gamma_{p+1}=\Pi^{a_1}\cdots\Pi^{a_{p+1}}\Gamma_{a_1\cdots a_{p+1}}$
is the static gauge worldvolume chirality operator.
Such a linear supersymmetry appears in the Green-Schwarz string
\ref\hupo{J. Hughes and J. Polchinski, \np{278}{1986}{147}.}; 
a general argument is given in \ref\linsusy{A. Achucarro, J. Gauntlett,
K. Itoh, and P.K. Townsend, \np{314}{1989}{129}.}.  
On the static gauge M-brane, we indeed see
the linear supersymmetry transformations generated by \supch, 
\eqn\lins{\eqalign{
\delta K^\mu=&~\bar\eta \Gamma^\mu\ferm\cr
\delta \phi^a=&~\bar\eta \Gamma^a\ferm\cr
\delta\ferm=&~(F_{\mu\nu}\Gamma^{\mu\nu} + \d_\mu\phi^a
	\Gamma^\mu\Gamma^a)\eta\ .\cr
}}
These transformations receive nonlinear corrections at higher order 
in field strengths (see \linsusy).

Since we do not at the moment possess a fully 
covariant formalism for the theory,
and sigma models with Ramond backgrounds are not as well understood as
ones with Neveu-Schwarz backgrounds, 
we haven't analyzed the fermionic terms to all orders in the fields. 
A cursory examination of the sort of four-point contact terms
that must arise indicates that the extension of the Goldstone
fields to their superspace counterparts is indeed occurring.

Another clue to the structure of the fermionic terms comes from the
expected relation between the null reduction to 1+1 dimensions
and the type IIB string \refs{\km,\kmo}.  
Since the M-brane world-volume theory
contains a Born-Infeld gauge field, we should look for a 
Dirac-Born-Infeld formulation of the Green-Schwarz string.
Fortunately, this has been worked out by Bergshoeff, London, and Townsend
\ref\blt{E. Bergshoeff, L.A.J. London, and P.K. Townsend,
hep-th/9206026; \cqg{9}{1992}{2545}; the core idea first appears in
P.K. Townsend, \pl{277}{1992}{285}.} (BLT).
The action takes the form
\eqn\bltact{
  \LL_{{\sst\rm IIB~str}}= 
	e^{-\Phi}\det[\Pi_i^a\Pi_j^a+2F_{ij}]
}
in terms of the superspace forms
\eqn\susyforms{\eqalign{
  \Pi^a=&~ d\phi^a - \fermbar^r\Gamma^a d\ferm^r \cr
  F    =&~ d A  +[\coeff{1}2d\phi^a\fermbar^r\Gamma^a
	\tau^3_{rs}d\ferm^s-\coeff14(\fermbar\Gamma^a d\ferm)
	(\fermbar\Gamma^a\tau^3 d\ferm)]\ . \cr
}}
Here $r,s=1,2$ labels the two like-chirality (in spacetime) spinors
of IIB supersymmetry, and $\tau^3$ is the usual Pauli matrix.
Solving the equation of motion for $A$ (which is not dynamical
in 2d) gives
\eqn\moreblt{
  \LL=e^{-\Phi}\det[\Pi_i^a\Pi_j^a] - T^2e^{\Phi} + T(F - dA)\ ;
}
eliminating the auxiliary field $\Phi$, 
one finds the conventional Green-Schwarz string action;
\moreblt\ is a superspace version of \bltone, restricted to
two dimensions.

An interesting feature of the action \bltact,
to which we shall return below,
is that it is spacetime scale invariant under
the transformations $\phi\rightarrow\lam\phi$, 
$\ferm\rightarrow\lam^{1/2}\ferm$,
$A\rightarrow\lam^2 A$, and $e^{\Phi}\rightarrow\lam^4 e^{\Phi}$
\blt.  The string tension $T$ arises from the expectation value of $F$,
spontaneously breaking this symmetry.

Since the M-brane action is only known so far in static gauge,
we should compare it to the action \moreblt\ gauged fixed via
$\d_i\phi^a=\delta_i^{\;a}$, and
$\Gamma_2\ferm^r=\tau^3_{rs}\ferm^s$ (for the definition of $\Gamma_2$,
see the discussion following \kappasym).
The superspace Wess-Zumino term of the Green-Schwarz string
action has a cubic part
\eqn\cubicwz{
  \LL_3^{{\sst WZ}}=T\eps^{ij}\d_i\phi^a\fermbar\Gamma^a\d_j\ferm\ .
}
In \moreblt, this expression appears in the quantity $F-dA$; in \lfull\
and \ltf,
it appears when expanding about a general complex structure.
Note its similarity to \twodwz; this means that instanton amplitudes
of the N=2 string are a source of the 2d Green-Schwarz WZ term of the
target worldsheet.  Thus one might obtain this term from
the superspace extension of \lfull\ \`a la BLT, or alternatively
from the complexification of \bltact;
however, we do not yet know what the
precise form of the action is that results in vanishing four-dimensional
S-matrix as required by N=2 string dynamics.  

The vanishing of the (2,1) string
S-matrix together with the known `volume term' \lfull\ in principle
determines the Wess-Zumino terms inductively, but the algebra
looks complicated.  Clearly a better
geometrical understanding of the relation between the Rarita-Schwinger
potential $\ferm$ and the sorts of `free differential algebras'
appearing in the dynamics of $p$-branes
\nref\fda{J. Azc\'arraga, J. Gauntlett, J.M. Izquierdo, 
and P.K. Townsend, \prl{63}{1989}{2443}.}%
\refs{\fda,\blt}
would be helpful.
The gauge corrections to the gravitational action \gract\ were
quite simply derived from the sigma-model anomaly.
This is an anomaly in $A\rightarrow A+d\Lambda$; in
the self-dual case, one has $\phi\rightarrow\phi+\Lambda$
at the linearized level, so the anomaly is connected
to spontaneous breaking of translation symmetry by the extended
object in our reinterpretation of the geometry.  It is quite plausible
that a similar approach will give the fermionic terms via a 
contribution of the $\vartheta$'s to the sigma-model anomaly,
this time for local supersymmetry transformations in spacetime
(\ie\ \stsusy\ in the self-dual case).
One needs to understand properly the spacetime supersymmetry
current algebra on the (2,1) string worldsheet.
It may be that some variant of the 
Green-Schwarz-Berkovits sigma model 
\ref\gsb{N. Berkovits, hep-th/9604123, and references therein.
N. Berkovits and W. Siegel, hep-th/9510106;
\np{462}{1996}{213}.
J. deBoer and K. Skenderis, hep-th/9608078.},
which manifests more of the spacetime supersymmetry, might allow one to 
determine this structure more easily.

It is important to complete the determination of the fermionic
terms in the M-brane worldvolume action; in particular, the Wess-Zumino
terms should be the flat spacetime expectation values of superspace
antisymmetric tensor fields, hence could help determine what sort
of fields couple to the M-brane.  Moreover, these terms will help
in uncovering the $\kappa$-symmetry of a more covariant formulation
of the M-brane worldvolume dynamics.  

\newsec{Discussion}

The main result of this paper is the derivation of the exact
target space action for the bosonic excitations of N=(2,1)
heterotic strings in uncompactified space-time, $\IR^{2,2}$,
given in equations \lfull, \nonab.  The form of this action,
and in particular its relation to the Nambu-Goto/Dirac-Born-Infeld
action for strings, provides new evidence for the
proposal of \refs{\green,\km,\kmo} that the target
space dynamics of heterotic N=2 strings in their two dimensional
vacua describes critical string worldsheets.
It also opens the way for a more thorough investigation of the
unification of string worldsheet and membrane worldvolume dynamics
in the framework of self-dual gauge theory coupled to self-dual gravity
in 2+2 dimensions, as described by N=2 heterotic strings. 

In the course of the discussion, we have encountered a few
interesting features of strings which our construction points to,
and are worth summarizing briefly:

\item{1)} The string target space dynamics seems to
mirror rather closely (and in some cases exactly) the worldsheet
structure\foot{Similar behavior appears to be exhibited
by certain topological string theories
\ref\efr{S. Elitzur, A. Forge and E. Rabinovici,
\np{388}{1992}{131}.}.}. 
This may have important consequences for understanding
target space dynamics in general, 
in particular the necessity to include
target spaces of different topologies, the cosmological constant
problem, etc.

\item{2)} In some of our constructions, the N=2 string coupling constant
appears to be fixed by non-perturbative N=2 string considerations.
This may be of interest to the question of vacuum selection
in more realistic string theories.

\item{3)} The target space strings one gets from N=2 heterotic string
are fixed at particularly symmetric points in the moduli spaces
of vacua.  This too may hint at the mechanism for vacuum selection
in string theory.  For instance, the initial state of the universe
near the big bang might involve such an exceptional
vacuum of string theory, compactified down to spatial dimensions of order
the Planck scale and possessing a very stringy sort of enhanced
symmetry.%

\item{4)} The target space dynamics of N=2 heterotic strings 
possesses unbroken symmetries that act nonlinearly on the
vertex operators (the small oscillations of target space fields),
and lead to crucial dynamical effects.
It is important to understand whether there are such symmetries
in more realistic string theories, and what is their role in the
dynamics.

\noindent
In summary, N=(2,1) heterotic strings provide a simple model
in which one might be able to address important issues in
string dynamics in a controlled setting.  

The 2+2d M-brane, as given to us by the target space
dynamics of N=2 heterotic strings, seems to contain 
the appropriate degrees of freedom needed to provide 
a unified presentation of all classes of string vacua.  
In addition, one discovers
a window into aspects of string theory that are
not seen by other probes -- hidden dimensions, hidden symmetries
such as spacetime conformal invariance, 
and underlying structure on the worldsheet,
to name a few.  These may be features of string theory
that are hard to detect except at the very symmetric vacua
forced upon us by self-consistency of the N=(2,1) string.
Let us discuss a few of them now.

\subsec{Null reduction and conformal symmetry}

The geometrical role of the null reduction \tj, \phst\ has been a
persistent puzzle in our construction.  Is it a feature of spacetime
or of the worldsheet?  Is O(10,2) symmetry restored in some
dynamical regime?  Recently, B.E.W. Nilsson suggested to us
\ref\tfebn{B.E.W. Nilsson, private communication.} 
that null reduction is related to a conformally covariant formulation
\ref\marnil{R. Marnelius and B.E.W. Nilsson, \prd{22}{1980}{830}.}%
\nref\emil{E. Martinec, hep-th/9608017.}%
\foot{In previous work
\refs{\km,\kmo,\emil}, we speculated that the extension of
O(9,1) to O(10,2) is related to conformal symmetry; the present
suggestion provides a concrete mechanism.}.
Dynamics of a point particle in ($d$--1)+1 dimensions
can be reformulated in a $d$+2 dimensional spacetime which linearly
realizes the conformal group O($d$,2).  To maintain equivalence to the
usual particle dynamics, one needs additional first-class
constraints:
\eqn\bnull{\eqalign{
  p^2=&~0\cr
  x\cdot p=&~0\cr
  x^2=&~0\ .
}}
The first of these is the usual mass-shell condition;
the second constraint forces dynamics to be orthogonal to the 
$d$+2 null cone selected by the third constraint
(for details of the gauge fixing, see \marnil).  Together 
these constraints form an Sp(2) algebra in phase space.
The symmetry under $x\leftrightarrow p$ is reminiscent of
T-duality.  The incorporation of spin follows similarly \marnil;
one adds a set of worldline fermions $\psi$ to represent
the algebra of Dirac matrices, together with the constraints
\eqn\fnull{\eqalign{
  p\cdot\psi=&~0\cr
  x\cdot\psi=&~0\ ,
}}
enlarging the algebra to OSp(2$|$1).  The first constraint is the
Dirac equation; the second reduces the spinor content from
$d$+2 to $d$ dimensions.  Vector \marnil\ and higher
antisymmetric tensor fields \ref\siegel{W. Siegel,
\mpl{A3}{1988}{2713}.}
are similarly incorporated (for instance, the Lorentz gauge
$\d\cdot A=0$ is augmented by $x\cdot A=0$).  Mass is represented
in this framework as momentum in the hidden dimensions
(the eigenvalue of the dilation operator).

The null reduction in N=2 heterotic string theory is remarkably similar
to \bnull, \fnull; the main difference being that $x^2=0$ has 
been eliminated, perhaps by gauge fixing.
What then remains are the Virasoro and super-Virasoro
constraints $P^2=0$, $P\cdot\psi=0$, and the null current
and supercurrent conditions $v\cdot P=0$, $v\cdot\psi=0$.

This connection between null reduction and conformal invariance 
resonates with the results of Bergshoeff \etal\
described in section 3 (equation \bltact\ and subsequent discussion),
where the breaking of spacetime scale invariance is related
to the expectation value of the Born-Infeld vector field strength.
It also might provide a deeper explanation for the strong
resemblance of the dilaton to a conformal compensator
in all critical string theories, as well as the close
connection between conformal and Poincare supergravity theories.
In \bnull\ \marnil, the mass scale is momentum in the hidden dimensions;
in \bltact\ \blt, it is the expectation value of the Born-Infeld vector
field; are the two related?
We will argue shortly that they are indeed.
It remains an open problem to formulate M-brane dynamics
in a way that respects the above OSp(2$|$1) symmetry,
as well as $\kappa$-symmetry and reparametrization invariance.

\subsec{Covariant formulation}

What are the ingredients required for a covariant formulation
of M-brane dynamics?
N=(2,1) heterotic strings present the M-brane in a fixed (static)
gauge, describing only the dynamics of the physical degrees
of freedom.  These consist of eight bosons and eight fermions on-shell.
A covariant formulation requires reintroduction of the four longitudinal
bosonic degrees of freedom eliminated via reparametrization
gauge fixing.
There are sixteen Green-Schwarz fermion fields appearing in the
physical gauge M-brane dynamics 
(the Dirac equation reducing this to eight on-shell);
covariant type II branes have 32 (two spinors of O(9,1)), 
compensated by a pair of eight-component local fermionic 
`$\kappa$-symmetries'.  A further doubling to 64 is needed
to have O(10,2) covariance before the null reduction.

Sixty-four component spinors are also the minimum required
such that the supersymmetry algebra realizes either IIA or IIB
supersymmetry with the same degrees of freedom, projected
in different ways \refs{\kmo,\emil}.  In other words,
one null reduction of the covariant formalism should yield
the D2-brane and IIA supersymmetry; another should yield 
the D-string and IIB supersymmetry.

Such a supersymmetry algebra has been investigated by Bars 
\ref\bars{I. Bars, hep-th/9607112.}
(for earlier work, see for example \ref\vhvp{J.W. van Holten
and A. van Proeyen, J.Phys. {\bf A15} (1982) 3763.}).
In fact, the starting point of \bars\ is in some sense
`11+2 dimensional' in order to have manifest eleven-dimensional 
Lorentz covariance under null reduction to IIA/M-theory;
IIB supersymmetry results from a
9+1/2+1 splitting of the algebra.  
An 11+2 superalgebra is not explicitly written;
only its various type IIA/type IIB projections are discussed.
These are
\eqn\typeAB{\eqalign{
  IIA:\qquad &~\{Q_\alpha,Q_\beta\}=
	(C\Gamma^m)_{\alpha\beta}P_m + (C\Gamma^{m_1m_2})_{\alpha\beta}
	Z_{m_1m_2}+(C\Gamma^{m_1\cdots m_5})_{\alpha\beta}
	Z_{m_1\cdots m_5}\cr
  IIB:\qquad &~\{Q_\alphabar^\abar,Q_\betabar^\bbar\}=
	\gamma^\mbar_{\alphabar\betabar}(c\tau_i)^{\abar\bbar}
		Z_\mbar^{(i)}
	+\gamma^{\mbar_1\cdots\mbar_3}c^{\abar\bbar}
		Z_{\mbar_1\cdots\mbar_3}
	+\gamma^{\mbar_1\cdots\mbar_5}(c\tau_i)^{\abar\bbar}
                Z_{\mbar_1\cdots\mbar_5}^{(i)}\ .
}}
Here $\Gamma$, $C$ are the 32$\times$32 Dirac matrices and charge
conjugation matrix for eleven dimensional supersymmetry; 
$\gamma$ are the chirally projected Dirac matrices relevant to
ten dimensional IIB supersymmetry; and
$\tau_i$ are Pauli matrices, and $c$ the charge conjugation
matrix, acting on the flavor indices $\abar,\bbar=1,2$
(the spinor indices of the `hidden dimensions';
see \bars\ for details).  
The $Z_{m_1\cdots m_p}$ are the various $p$-form central charges
carried by the states of the theory.  
In the IIB reduction,
the momentum is $P_{\mbar}=Z^{(2)}_\mbar$; the remaining one-form
charges $Z^{(1)}_\mbar$, $Z^{(3)}_\mbar$ couple to the NS and RR
$B$-fields of the IIB F- and D-strings, respectively.
The algebra appears to admit an O(2,1) symmetry on the
indices $i,\abar,\bbar$; in string theory one only sees the $O(2)$
that preserves the momentum $P_\mbar$, and even this is broken to the
${\bf Z}_2$ of electric-magnetic duality 
$({\ 0\;\  1\atop -1\; 0})$ 
inside the SL(2,Z) S-duality symmetry of type IIB.

To facilitate comparison
to our work, for the remainder of this subsection
let us adopt the conventions and notation of \bars:
Timelike indices are labelled 0,$0'$; spacelike indices $1,2,\ldots,11$.
As an ansatz, identify coordinates $0',0,1,\dots9,11$ as the 10+2
spacetime dimensions of the M-brane formalism, and coordinate
10 as part of the field space of the Born-Infeld vector field.
Orient a 2+2 brane in the hyperplane spanned by 0,$0'$,9,11;
let the space transverse to the brane be spanned by indices 
$1,2,\ldots,8$.  This leaves the 10 direction, which carries the
IIA interpretation as the tenth spatial coordinate of 
`visible' spacetime, related to the Born-Infeld vector field by a 
duality transformation on the two-brane worldvolume.  
The zero mode of $F$ is momentum/winding in this extra
dimension via $^*F=d\phi$.
We cannot have such an interpretation in the IIB theory, where the
gauge field has no dual -- only its flux is a physical degree
of freedom.  In order to maintain a uniform language 
for both IIA and IIB, we drop
manifest 10+1 Lorentz symmetry in M-theory
and use the vector field description,
remembering that its flux represents motion/wrapping in an eleventh
dimension which is sometimes physical and sometimes a gauge coordinate
apart from its zero-mode\foot{One might be able to regard this extra
coordinate as the fiber of the line bundle whose connection
is the Born-Infeld vector potential $K$, much as the transverse
directions to the brane comprise the bundle $\EE$
described in the introduction.}.  
Thus this `dimension' (the one with index 10)
is real or hidden depending on the reduction.
In what follows, we always associate the 10 `direction'
with the flux of the Born-Infeld vector field.

With these conventions, the D2-brane of the IIA
theory results from null reduction along (for example) the $0'$-8 plane;
the two-brane lies along $0,9,11$, with transverse space $1,2,\ldots,7$
(and one more from the vector field, \ie\ the 10 direction):
\eqn\IIA{
  \left({0'\atop B}\right){0\atop B}
	\;\biggl|\;{1\atop T}{2\atop T}
	{3\atop T} {4\atop T}{5\atop T}{6\atop T}{7\atop T}
	\left({8\atop T}\right){9\atop B}{10\atop F}{11\atop B}\quad.
}
Here $B$ denotes a brane direction, $T$ a transverse direction,
and $F$ the flux direction; parenthesis denote directions
spanned by the null vector.
On the other hand,
the D1-brane of IIB results from splitting $0',10,11$ from 
$0,1,\ldots,9$, and regarding the former as
`hidden dimensions'.  In the M-brane formalism, two of these three
hidden IIB directions ($0'$,11) are eliminated by the null reduction;
the 10 coordinate is eliminated (apart from its zero mode)
by the gauge symmetry of the Born-Infeld vector field.  
With the null reduction along the $0'$-11 plane,
the D-string is along $0,9$ with transverse space $1,2,\ldots,8$:
\eqn\IIB{
  \left({0'\atop B}\right){0\atop B}\;\biggl|\;{1\atop T}{2\atop T}
	{3\atop T} {4\atop T}{5\atop T}{6\atop T}{7\atop T}
	{8\atop T}{9\atop B}\left[{10\atop F}\right]
	\left({11\atop B}\right)\quad.
}
The square brackets denote the `hidden' nature of the 10 direction.
As mentioned above, 
in the IIB algebra \typeAB, the SO(2) that rotates the IIB string
into the D-string is rotation in the 10-11 plane.
Both of these directions are hidden (\ie\ gauge coordinates)
in the M-brane; one corresponds to the spatial coordinate of the
M-brane worldvolume eliminated by the null reduction, the other
to the Born-Infeld vector field on the effective D-string
(for example, its spatial component in $A_0=0$ gauge).  
Just as 10+1d Lorentz symmetry is
non-manifest in the IIA reduction, the SL(2,Z) S-duality of
the type IIB reduction is non-manifest as well.  Nevertheless,
if the above identification is correct, we have found the relation
of our framework to F-theory \vafa.
Note that the two hidden dimensions of F-theory are indeed
spacelike here, and {\it not the same} as the extra 1+1
dimensions introduced via null reduction (one spatial coordinate
is common to both).  The identification of the Born-Infeld
flux with one of the coordinates of the F-theory torus 
(whose modulus is the complex IIB coupling constant)
is consistent with the fact that the electric charge 
of $(p,q)$ strings is represented on the D-string by this flux 
\ref\wittenpqstr{E. Witten, hep-th/9510135;
\np{460}{1996}{335}.},
while in F-theory this charge is supposed to be represented by the
winding number of some brane (conjectured to be the
self-dual IIB 3-brane) along the A-cycle of the hidden torus.
In the work of BLT described in section 3, 
and in the formula for the string tension of $(p,q)$ strings 
\nref\schwarz{J. Schwarz, hep-th/9508143; \pl{360}{1995}{13}.}%
\nref\schmidhuber{C. Schmidhuber, hep-th/9601003; \np{467}{1996}{146}.}%
\refs{\wittenpqstr,\schwarz,\schmidhuber},
flux of the Born-Infeld
vector field generates a shift of the string tension.
Here we have regarded it as momentum/winding in a `hidden
dimension', very much in concert with the remarks of the
previous subsection.
In other words, in particle dynamics mass is momentum in the
hidden directions, spontaneously breaking scale transformations
in O(d,2); in string theory the analog is the string tension,
which might again be interpreted as momentum/winding in hidden dimensions.
Finally, note that we see hints of extra coordinates, and non-manifest 
symmetries beyond O(10,2), as advocated in \emil.

\subsec{{\it Dynamics in the compactified case}}

When one (or more) of the spatial directions on $\IR^{2,2}$
is compactified, the theory starts probing the chiral,
left-moving CFT that, on $\IR^{2,2}$, is frozen by the
physical state conditions. This means, as discussed in
section {\it 2.4}, that oscillations in these ``internal''
directions can be excited and lead to the exponential density
of states described after equation \mn. In addition, the
effective dimensionality of the target space increases;
to see that, one notes that it is possible to generalize the
states \vb, \gfone\ to:
\eqn\genz{
  V_\xi=\int d\theta\int d^2\bar\theta\;
	\xi_\mu\psi^\mu e^{ik\cdot x+i\bar k\cdot\bar x}
}
where $\xi$, $k$ are 10+2 dimensional vectors
satisfying the usual physical state and null reduction
conditions, and the level matching condition $k^2=\bar k^2$
now has solutions for arbitrary momenta in the eight
left-moving directions. Thus, the theory becomes effectively
ten dimensional!
Of course, the additional allowed momenta lie on the $E_8$
torus, supporting the picture advocated in section 2 -- the
internal space of the (2,1) string mirrors the spacetime geometry.

Consider for concreteness the type IIB ``target worldsheet''
construction of Example 3 in section {\it 2.2}. Take the IIB
worldsheet to be parametrized by $(x^1, x^2)$ (i.e. gauge
$J=\partial x^0+\partial x^3$); compactify $x^2$ on a circle
of radius $R$. Denoting the momentum in the $(x^4, \cdots, x^{11})$
directions (which lies in the $E_8$ root lattice) by $\vec p$,
and recalling that the left and right moving momenta in the compact
$x^2$ direction are $k_2={n\over R}-{mR\over2}$; 
$\bar k_2={n\over R}+{mR\over2}$, level matching implies:
\eqn\levmat{(\vec p)^2=2nm.}
we see that the vertex operators $V_\xi$ describe a $9+1$
dimensional $U(1)$ gauge field on a compact space (the
$E_8$ torus $\times $ $S^1$, with the extra peculiarity
that the $S^1$ momentum $k^2$ actually corresponds to
mixed momentum and winding). Of course, in addition to
this gauge field the spectrum also includes the excited
states \mn.

The fact that compactification leads to a higher dimensional
``underlying'' theory is reminiscent of a similar phenomenon in \bfss\
where compactification of M-theory on a $p$-torus
is described in terms of a $p+1$ dimensional worldvolume
(related by T-duality to zero-brane quantum mechanics).

The dynamics of the 9+1 dimensional ``gauge field''
\genz\ and its massive relatives \mn\ provides a
highly non-trivial generalization of the dynamics studied
earlier in section 3.  As mentioned above, the only non-vanishing
S-matrix elements are still the three-point couplings. There
is presumably a non-polynomial generalization of  
\lfull, \nonab, which would be very interesting to find.
To leading order, the dynamics of the gauge field \genz\ is described
by a 10+2 dimensional supersymmetric gauge theory with a null 
reduction to 9+1d. Such a theory has been recently studied in 
\ref\sezg{H. Nishino and E. Sezgin, hep-th/9607185.}.

As discussed in section {\it 2.4}, compactified N=2
heterotic strings exhibit similarities to some properties
of BPS states and string duality. We next comment further
on these relations, starting with properties of BPS states.

Consider the heterotic string toroidally compactified
on an even self-dual Narain torus, $\Gamma^{24,8}$. 
{}From our discussion of N=2 strings we expect a special role
to be played by the compactification for which $\Gamma^{24,8}$
decomposes into separate even self-dual tori for the
left and right movers, $\Gamma^{24}$, and $\Gamma^8$ respectively.
The former is a Niemeier torus; the latter, the $E_8$ torus.

It is well known \dh\ that this system has an infinite number
of BPS states of arbitarily high mass, for which the (N=1
superconformal) right movers are in their ground state
(with momentum $\vec p_r\in \Gamma^8$), while the
(bosonic) left movers are excited at level $N$ (and with
$\vec p_l\in\Gamma^{24}$), such that:
\eqn\dabhar{M^2=(\vec p_r)^2=(\vec p_l)^2+2(N-1).}
One can focus on the physics of BPS states by noting that
the right moving ground state in \dabhar\ in fact preserves
N=2 worldsheet supersymmetry; gauging the global right moving
N=2 superconformal symmetry leaves the BPS states while projecting
out all non-BPS ones. In order for the resulting N=(2,0)
string to be critical, one has to do two things:

\item{1)} Add two more (non-compact) dimensions to make the
non-compact system 2+2 dimensional. Of course, the introduction
of the additional coordinates is harmless, since they disappear
upon the left moving null gauging discussed in section~2.

\item{2)} Remove the right moving $\hat c=8$ system corresponding
to the $E_8$ torus to make the N=2 string critical.
One way to accomplish this is by a chiral topological twist  
\ref\bgr{L. Baulieu, M. Green, and 
E. Rabinovici, hep-th/9611136.}.

\noindent
The latter condition means that only states with $\vec p_r=0$ survive
the gauging. Clearly, from \dabhar, there is a finite number of such
states, however upon further compactification one finds a large
subset of the BPS states, namely those with vanishing charges
in the $\Gamma^8$ directions but arbitrary left moving excitation.

The N=(2,0) heterotic string can be thought of describing the
physics of those states\foot{The remarks in the remainder of
this subsection derive from discussions with G. Moore.}. 
In particular, the algebra of BPS
states \harvmooretwo\ can be thought of (for the appropriate
subset of BPS states) as the OPE algebra of physical N=(2,0)
string states.  
In other words, the BPS algebra is merely the
three-point S-matrix of the N=2 string, its only nontrivial
scattering amplitude!
The kinematic condition that two BPS states make another on-shell
amounts in the N=2 string to the requirement that their momenta lie
in the same self-dual null plane.
The vanishing of the S-matrix for the tower of states \genz\
should be related to a symmetry group, much larger
than the usual symmetries of self-dual gravity and
self-dual Yang-Mills (discussed for example in \refs{\ovone,\lmns,\popov}).
It is tempting to speculate, along the lines of 
\nref\givshap{
A. Giveon and M. Porrati, \np{355}{1991}{422};
A. Giveon and A. Shapere, hep-th/9203008;
\np{386}{1992}{43}.}%
\nref\gregfinite{G. Moore, hep-th/9305139.}%
\refs{\givshap,\gregfinite},
that these S-matrices are related to the structure constants
of a large broken symmetry group of spacetime
(\eg\ for the N=(2,0) string, self-dual Yang-Mills with gauge algebra
the Fake Monster Lie Algebra).
In this connection, it is interesting that
denominator formulae corresponding to generalized
Kac-Moody algebras may be obtained by evaluating the one loop
partition sums of various N=(2,0) strings, using formulae of
\harvmooretwo.  Consider the N=(2,0) string with internal sector
compactified on the Leech torus, and four-dimensional spacetime
$\IR^2\times T^2$.  It was pointed out in
\km\ (section {\it 6.1})
that the one-loop partition function is
\nref\jfunction{G. Lopes Cardoso, D. L\"ust, and T. Mohaupt,
\np{450}{1995}{115}; hep-th/9412209.}%
\refs{\jfunction,\harvmooretwo}
\eqn\hmres{\eqalign{
  Z_1(T,U)=&~
	{1\over2}\int_\FF {d^2\tau\over\tau_2 }
        	\biggl(\;\sum_{(p_L, p_R)}
		q^{{1\over2}p_L^2}\bar q^{{1\over2}p_R^2}
		\biggr)\bigl(J(\tau)+24\bigr)\cr
	=&~ -24\log\left(\sqrt{T_2}|\eta(T)|^2
        	\sqrt{U_2}|\eta(U)|^2\right)-\log|J(T)-J(U)|^2\ .\cr
}}
Here $T$, $U$ are the complex structure and \kahler\
moduli of the $T^2$; $J(\tau)$ is the unique modular form with both
a simple pole of unit residue and no constant term 
in its $q$-expansion\foot{The 24 here arises from the oscillator
states at level one, the only massless states in a compactification
on the Leech torus.  Compactification on some other Niemeier torus
would replace this number by $C_0$, the dimension of the Niemeier
group.}.
This particular compactification is a
special point in the Narain moduli space $\Gamma^{2,26}$
of the N=(2,0) string.  From \ref\borcherds{R. Borcherds,
{Invent. Math.} {\bf 120} (1995) 161.}, \harvmooretwo,
we know that this result extends uniquely to an automorphic form
over the full Narain moduli space; this automorphic form is
the denominator formula for the so-called Fake Monster Lie
Algebra.  Repeating the same excercise with the
Monster module as the internal space
yields the denominator product for the Monster Lie Algebra.

A similar relation exists between the sector
of BPS states of type II superstrings compactified on $\Gamma^{8,8}
=\Gamma^8\oplus\Gamma^8$ and the N=(2,1) string.
The (2,1) string on the spatial Narain torus $\Gamma^{2,10}$
involves the appropriate GKM superalgebra related to $II^{1,9}$.
If one factors out the trivial zero due to boson-fermion
cancellation, one should obtain a denominator product for
this superalgebra.

In the context of string duality, an interesting question
concerns the relation of winding N=2 strings and dyonic 
strings (see \dvv\ for a related discussion). To discuss this,
we return to Example 3 of section 2.2. If the spatial direction
of the IIB worldsheet, $x^2$, is compact, the eight bosonic and 
fermionic target space massless fields describing the IIB
worldsheet acquire ``magnetic'' partners obtained by exchanging
momentum and winding on the N=2 string. In addition we find
a tower of mixed momentum winding states \mn. It is natural to
interpret the pure winding N=2 string states as describing a
``magnetic'' type IIB string, and the mixed momentum/winding
ones as dyonic type IIB strings. Such an interpretation would 
provide a link between the physics of BPS states in space-time
and string duality, and in particular explain the appearance of
GKM algebras in both contexts.

\subsec{Relation to other work}

{\sl (a) Self-duality}

As was mentioned in section 3, N=2 worldsheet
supersymmetry in two complex target
dimensions automatically extends to N=4 supersymmetry,
with the additional currents built by U(1) spectral flow
(see the discussion before equation \iij).  Thus backgrounds
of N=2 strings admit a triplet of complex structures \iij;
N=2 heterotic strings possess a torsion background $H$
that obstructs closure, $dI=H$ (\cf\ \oneloop), and hence the
hyper-complex geometry is not \kahler\ (although by abuse of language 
such geometries have been called `hyper-\kahler\ with torsion').

The field-theoretic action \sfourwz\ that describes
the M-brane in the absence of self-dual gravity
has been studied by Nair and Schiff \dns,
and by Losev, Moore, Nekrasov and Shatashvili (LMNS) \lmns.  
This action is a natural
lifting of the two-dimensional WZW theory to four dimensions;
indeed, we have seen that null reduction of \sfourwz\ reduces
to the 2d chiral model (with a WZ term, depending on the relative
orientation of the null reduction and the complex structure).
Nair and Schiff and LMNS have shown that much of 
rational conformal field theory
has a counterpart in higher dimensions.
Many two-dimensional features lift to four dimensions -- the
Polyakov-Wiegmann identity, holomorphic
anomalies, holomorphic factorization, current algebra Ward identities,
$b$-$c$ systems, and so on.
Symmetry considerations determine
the form of an `algebraic sector' of the theory (a certain
subset of correlations \lmns), although these do not provide enough
data to reconstruct the full theory as is the case 
in two dimensions.
Roughly speaking, the holomorphic gauge transformations one uses to
derive the Ward identities depend on functions of two variables
($x^i$, $i=1,2$), whereas the boundary data specifying the
dynamics depend on functions of three variables.
In the case at hand, however,
one is interested in the dimensional reduction of the dynamics to
3d or 2d, in which case the algebraic sector -- with a
few global considerations -- might be sufficient.

LMNS have derived the target space action of open and closed
N=(2,2) strings using the holomorphic anomaly of free fields
in the presence of a background gauge and gravitational connection.
The (2,2) string was obtained because the background geometry was
chosen to be \kahler; one might hope that an extension to the 
hyper-complex case will generate the action \lfull.
One might look for an analog of the equivalence between
Nambu-Goto and Polyakov actions that exists in the two-dimensional
quantum theory.  The work of \lmns\ shows this to be the case
for the algebraic sector of the target space field theory
of the N=(2,2) string.  One caveat to this approach is 
the above-mentioned incompleteness of the algebraic sector; 
there might not be complete equivalence between the free-field dynamics
and that of the effective gauge action \sfourwz.
However, one property that one might be able to extract from the results of
LMNS, appropriately extended to Hermitian metrics, is a determination
of the critical dimension directly from anomalies in the M-brane 
field theory.  Clearly the N=(2,1) string gives a 12d spacetime
as the target space of the M-brane; one would like to see this
as a consistency condition for the quantization of reparametrization
and $\kappa$-symmetries, just as anomaly cancellation on the
string generates the critical dimension d=10.

Apart from considerations of quantization, there is a new and
unexpected integrability underlying membrane dynamics
in special backgrounds,
derived from self-duality.  A standard technique for generating
classical solutions to self-dual dynamics
is the twistor transform.  A wrinkle in
the application of twistor ideas to the present case is again the
presence of torsion in the geometry, which is involved in the 
integrability conditions on the triplet of complex structures.
This situation has been considered by Howe and Papadopoulos
\ref\hopa{P. Howe and G. Papadopoulos, hep-th/9602108;
\pl{379}{1996}{80}.},
who have given a construction of the twistor space for the
relevant hyper-complex geometry; it is again essentially
a fibration of the four-manifold over the $\IC{\bf P}^1$ of
complex structures \iij.  
It would be interesting
to apply the twistor transform to construct exact solutions
of the 2+2d and 2+1d dynamics of brane waves governed by
\lfull.

\medskip
{\sl (b) M-theory as a matrix model}

Banks \etal\ \bfss\ have proposed a definition of M-theory as a matrix
model. In the regime where the matrices are approximately commutative,
their eigenvalues describe the dynamics of zero-brane `partons'
of an infinite momentum frame formulation of eleven-dimensional
supergravity.  When the commutators are large, the theory
provides a regularized description of membranes.  How might this
proposal be related to the N=2 string?  

There are a number of similarities
between N=2 string theory and matrix models of (two dimensional)
noncritical strings.
Both have dynamics in only the center-of-mass degrees of freedom
of the string.  Both give rise to integrable field theories 
in target space.  The S-matrix is largely trivial.  
Area-preserving diffeomorphisms seem to play a role in each.
The noncritical string field theory and the matrix collective
field are related by a nontrivial transform similar
to a Backlund transformation; similarly, the integrable
theory \lfull\ is nontrivially related to the usual supermembrane
field theory.

The N=(2,1) string describes a particular state in M-theory --
a long, stretched membrane (in static gauge); N=(2,1) strings
are the quanta of small excitations about this state.
In a matrix model, the long stretched membrane would correspond to
a particular master field of the large N limit.
One might imagine that N=2 strings describe the perturbative quantization
(in powers of the inverse membrane tension) of the collective field
theory expanded about this master field, much as noncritical
strings encode the perturbative expansion about the
collective field of matrix quantum mechanics
(the scaling limit of the eigenvalue distribution).
It may well be that N=(2,1) strings also have a nonperturbative
ambiguity like that of noncritical strings, here related to
the fingering instability of membranes -- something that
would not be seen in the perturbation expansion about a long
stretched membrane.
It would also be interesting to see if, as one might expect
on the basis of the general arguments of Shenker
\ref\shenker{S. Shenker, in Carg\`ese 1990, {\it Random
Surfaces, Quantum Gravity, and Strings}, O. Alvarez \etal,
eds.; NATO ASI ser. 262.},
there are large $O(\exp[-1/g_{str}])$ nonperturbative effects
in the (2,1) string, which in this interpretation might be
zero-brane/eigenvalue tunnelling processes.


\bigskip
\noindent{\bf Acknowledgements:}  We thank
N. Berkovits,
M. Cederwall,
J. de Boer,
R. Dijkgraaf,
M. Green,
J. Harvey,
C. Hull,
E. Rabinovici,
A. Schwimmer,
S. Shatashvili,
K. Skenderis,
P. Townsend,
E. Verlinde,
and
H. Verlinde
for discussions.
We are especially grateful to 
G. Moore and B. Nilsson for illuminating 
remarks and conversations, and to G. Moore
for comments on the manuscript.
Various parts of this work were carried out during
(and presented during)
the 1996 Duality Workshop at the Aspen Center for Physics 
(E.M., D.K.),
The Fourth Nordic Meeting on Supersymmetric Field
and String Theories at G\"oteborg University 
(E.M.), 
and the Four Dimensional Geometry and Quantum Field Theory
workshop at The Newton Institute for Mathematical Sciences 
(E.M., D.K.); 
we thank these institutions for
their hospitality.  A preliminary account of this work
was given by the authors at Strings '96, 
(Santa Barbara, July 15-20, 1996).

\bigskip
{\bf Note added:}
At the time of completion of the present work, there appeared
a paper of Hewson and Perry \ref\hewper{S. Hewson and M. Perry,
hep-th/9612008.} investigating a theory of 2+2 branes in
10+2 dimensions.  It would be interesting to understand the relation
of their work to ours.  In particular, their construction is a 
`$p$-brane', having no world-volume vector field, whereas ours
is more along the lines of the $D$-brane.


\appendix{A}{Null reductions and partition sums.}

The purpose of this Appendix is to comment on the
different null reductions to 1+1 and 2+1 dimensions,
and in particular on the question of modular invariance
of the corresponding N=2 worldsheet theories. 
For simplicity, we
discuss the vacua of the N=(2,0) string on Niemeier
tori described in Example 1 of section {\it 2.2}. 
Generalization to other cases is straightforward.

Since we are gauging a null (anomaly free) symmetry,
the only effect of the gauging on the N=2 heterotic
string path integral on a worldsheet torus, is
the imposition of the constraint $\delta(p\cdot v)$ 
on the momentum $p$ flowing around the torus.
This constraint can be thought of as arising from the 
integral over the zero mode of the gauge field which couples
to $J$ \tj. One might be worried that for the 2+1d
null reduction, the procedure could break modular
invariance, since it eliminates one of the twenty-four chiral
scalars $x^a$ \tj, and there are of course no
23 dimensional even self-dual lattices. To examine
this issue we study here in turn the partition sums
corresponding to the 1+1 and 2+1 dimensional cases. 

Consider first the null reduction \tj, \phst\ to
1+1d, achieved by gauging \eg:
\eqn\aaa{J=\partial x_1+\partial x_3.}
The effect of \aaa\ on the zero mode part of
the torus path integral is the replacement:
\eqn\bbb{\int dp_1 dp_3 \left(q\bar q\right)^{{1\over2}(p_3^2-p_1^2)}
\to \int dp_1 dp_3 \left(q\bar q\right)^{{1\over2}(p_3^2-p_1^2)}
\delta(p_3-p_1)}
Before the replacement, the zero mode integral over
$p_1$, $p_3$ gave (after an appropriate Wick rotation)
$1/\tau_2$. After imposing the constraint, we can perform
the integral over $p_1$, and are left with:
$\int dp_3$. States corresponding to different $p_3$ are
identified by spectral flow in the null U(1), 
and thus this infinite degeneracy should be discarded.
The net effect of the null reduction is to eliminate
the zero modes of $x_1$, $x_3$. It is easy to see that the
non-zero modes are eliminated as well.
Hence, the full partition sum of the N=(2,0) string whose
momenta live on a Niemeier
lattice $\Lambda$ is:
\eqn\ddd{Z(\tau)={1\over\eta^{24}(\tau)}
\sum_{p\in\Lambda}q^{{1\over2}p^2}.}
The spectrum of physical states one reads off from \ddd\
agrees with the analysis of section~2.

Reduction to 2+1d is achieved by gauging (say):
\eqn\ccc{J=\partial x_1+\vec v\cdot \vec\partial x.}
where $\vec v$ is a unit vector which
points along the Niemeier torus associated
with the twenty four chiral scalars $x^a$.
After imposing the delta function $\delta(p_1-\vec v\cdot
\vec p)$ in the path integral, and performing the 
integral over $p_1$, we now find the partition sum:
\eqn\eee{Z_0={1\over\eta^{24}(\tau)}
\sum_{p\in\Lambda} q^{{1\over2}[p^2-(\vec p\cdot \vec
v)^2]}\bar q^{-{1\over2}(\vec p\cdot \vec v)^2}}
The form in square brackets implies that one combination
of the twenty-four scalars $x^a$ decouples, and we are left with
an effectively 23 dimensional left moving lattice. However,
the $\bar q$ dependence indicates that the missing
combination reappears on the right moving side, with
negative norm. Thus, overall, we find that momenta live
on a 23+1 dimensional lattice where all twenty-four dimensions
have positive norm, the first twenty-three trivially, and the
twenty-fourth because of a cancellation of two minus signs.
One is due to its
right moving nature (as in Narain compactifications); the other
to its being timelike \eee.

Hence, the 2+1 dimensional vacuum leads to a modular
invariant theory; however, the partition sum \eee\ makes it clear
that the vacuum is unstable, due to a tachyonic divergence
arising from states with $\vec p\cdot \vec v\not=0$ (see the
discussion in the text following eq. \dispr).

Note that, as mentioned in the text, a potential
2+2 dimensional vacuum of N=2 strings is rendered
inconsistent by the above analysis. Indeed, if we
try to gauge, \eg,
\eqn\fff{J=\partial x_4+i\partial x_5\ ,}
the constraint $p_4+ip_5=0$ imposed on the
path integral implies that we are summing
over the sublattice of $\Lambda$ with 
$p_4=p_5=0$. This twenty-two dimensional lattice
does not lead to a modular invariant partition
sum, and hence corresponds to an inconsistent theory.
 
\bigskip

\appendix{B}{Comments on quantization.}

While (uncompactified) N=2 heterotic strings
have a field theoretic spectrum, their quantization 
might nevertheless be subtle. We have seen one
aspect of this in this paper: compactifying N=2
heterotic strings on a circle reveals a rich spectrum
of states absent from the target field theoretic description.
A related subtlety arises when one attempts to study the
system at finite temperature by Euclideanizing time and
making it compact.

In addition, it has been pointed out in the past (see \eg\
\ovone) that loop amplitudes in N=2 string theory
appear to differ from those one would write
down in the target space field theory with the same classical
spectrum and interactions. As an example, at one loop,
in addition
to the usual Schwinger parameter,
which corresponds to the imaginary part of the
modulus of the torus $\tau$, one finds in N=2 string theory
a second modulus $u$ related to the U(1) gauge field (see
section {\it 2.1}).  Since this U(1) modulus lives on
the Jacobian of the torus, the measure for integrating
over $u$ involves a factor of ${\it Im}\tau$, which affects
power-counting of momentum integrals \ovone.
Moreover, modular invariance cuts off ultraviolet divergences
in a very non field-theoretic way.

One can in fact see some of these issues classically.
Consider an N=2 string in a linear dilaton background,
$\Phi(x)=Q_\mu x^\mu= Q^i x^{\bar i}+Q^{\bar i} x^i$.
Criticality requires that $Q^2=Q\cdot \bar Q=0$. As is well
known, the linear dilaton modifies the superconformal
generators \ntwo\ to:
\eqn\tmod{\eqalign{
  \bar T=&~-{1\over2}\left(\bar\partial x\cdot 
  	\bar\partial x+Q\cdot \bar\partial^2 x +
	\bar\psi\cdot\bar\partial\bar\psi\right)\cr
  \bar G^\pm=&~(\eta_{\mu\nu}\pm I_{\mu\nu})
	\left(\bar\psi^\mu\bar\partial x^\nu
 	+Q^\mu\bar\partial\psi^\nu\right) \cr
  \bar J=&~\half I_{\mu\nu}\bar\psi^\mu\bar\psi^\nu+
	I_{\mu\nu}Q^\mu\bar\partial x^\nu~.  \cr}
}
The modified form of $\bar J$ leads to a change 
in the physical state conditions.
While before, physical states such as \vbos, \gf\ \etc\
had to satisfy the 
physical state condition $k^2=0$, now one finds 
{\it two} conditions which can be written as:
\eqn\bone{k_\mu(k^\mu+Q^\mu)=0~;\;\;\; k_\mu Q_\nu I^{\mu\nu}=0}
or in complex coordinates:
\eqn\btwo{k\cdot(\bar k+\bar Q)=0~; \;\;\;\bar k\cdot (k+Q)=0.}
The theory behaves as if the kinetic operator itself has been
complexified in going from 1+1 to 2+2d.
Target space field theory in a linear dilaton background
would reproduce naturally the first of equations \bone.
It remains to be understood how it knows about the second.

\listrefs
\end


+Q^\mu)=0~;\;\;\; k_\mu Q_\nu I^{\mu\nu}=0}
or in complex coordinates:
\eqn\btwo{k\cdot(\bar k+\bar Q)=0~; \;\;\;\bar k\cdot (k+Q)=0.}
The theory behaves as if the kinetic operator itself has been
complexified in going from 1+1 to 2+2d.
Target space field theory in a linear dilaton background
would reproduce naturally the first of equations \bone.
It remains to be understood how it knows about the second.

\listrefs
\end

ambiguity like that of noncritical strings, here related to
the fingering instability of membranes -- something that
would not be seen in the perturbation expansion about a long
stretched membrane.
It would also be interesting to see if, as one might expect
on the basis of the general arguments of Shenker
\ref\shenker{S. Shenker, in Carg\`ese 1990, {\it Random
Surfaces, Quantum Gravity, and Strings}, O. Alvarez \etal,
eds.; NATO ASI ser. 262.},
there are large $O(\exp[-1/g_{str}])$ nonperturbative effects
in the (2,1) string, which in this interpretation might be
zero-brane/eigenvalue tunnelling processes.


\bigskip
\noindent{\bf Acknowledgements:}  We thank
N. Berkovits,
M. Cederwall,
J. de Boer,
R. Dijkgraaf,
M. Green,
J. Harvey,
C. Hull,
E. Rabinovici,
A. Schwimmer,
S. Shatashvili,
K. Skenderis,
P. Townsend,
E. Verlinde,
and
H. Verlinde
for discussions.
We are especially grateful to 
G. Moore and B. Nilsson for illuminating 
remarks and conversations, and to G. Moore
for comments on the manuscript.
Various parts of this work were carried out during
(and presented during)
the 1996 Duality Workshop at the Aspen Center for Physics 
(E.M., D.K.),
The Fourth Nordic Meeting on Supersymmetric Field
and String Theories at G\"oteborg University 
(E.M.), 
and the Four Dimensional Geometry and Quantum Field Theory
workshop at The Newton Institute for Mathematical Sciences 
(E.M., D.K.); 
we thank these institutions for
their hospitality.  A preliminary account of this work
was given by the authors at Strings '96, 
(Santa Barbara, July 15-20, 1996).

\appendix{A}{Null reductions and partition sums.}

The purpose of this Appendix is to comment on the
different null reductions to 1+1 and 2+1 dimensions,
and in particular on the question of modular invariance
of the corresponding N=2 worldsheet theories. 
For simplicity, we
discuss the vacua of the N=(2,0) string on Niemeier
tori described in Example 1 of section {\it 2.2}. 
Generalization to other cases is straightforward.

Since we are gauging a null (anomaly free) symmetry,
the only effect of the gauging on the N=2 heterotic
string path integral on a worldsheet torus, is
the imposition of the constraint $\delta(p\cdot v)$ 
on the momentum $p$ flowing around the torus.
This constraint can be thought of as arising from the 
integral over the zero mode of the gauge field which couples
to $J$ \tj. One might be worried that for the 2+1d
null reduction, the procedure could break modular
invariance, since it eliminates one of the twenty-four chiral
scalars $x^a$ \tj, and there are of course no
23 dimensional even self-dual lattices. To examine
this issue we study here in turn the partition sums
corresponding to the 1+1 and 2+1 dimensional cases. 

Consider first the null reduction \tj, \phst\ to
1+1d, achieved by gauging \eg:
\eqn\aaa{J=\partial x_1+\partial x_3.}
The effect of \aaa\ on the zero mode part of
the torus path integral is the replacement:
\eqn\bbb{\int dp_1 dp_3 \left(q\bar q\right)^{{1\over2}(p_3^2-p_1^2)}
\to \int dp_1 dp_3 \left(q\bar q\right)^{{1\over2}(p_3^2-p_1^2)}
\delta(p_3-p_1)}
Before the replacement, the zero mode integral over
$p_1$, $p_3$ gave (after an appropriate Wick rotation)
$1/\tau_2$. After imposing the constraint, we can perform
the integral over $p_1$, and are left with:
$\int dp_3$. States corresponding to different $p_3$ are
identified by spectral flow in the null U(1), 
and thus this infinite degeneracy should be discarded.
The net effect of the null reduction is to eliminate
the zero modes of $x_1$, $x_3$. It is easy to see that the
non-zero modes are eliminated as well.
Hence, the full partition sum of the N=(2,0) string whose
momenta live on a Niemeier
lattice $\Lambda$ is:
\eqn\ddd{Z(\tau)={1\over\eta^{24}(\tau)}
\sum_{p\in\Lambda}q^{{1\over2}p^2}.}
The spectrum of physical states one reads off from \ddd\
agrees with the analysis of section~2.

Reduction to 2+1d is achieved by gauging (say):
\eqn\ccc{J=\partial x_1+\vec v\cdot \vec\partial x.}
where $\vec v$ is a unit vector which
points along the Niemeier torus associated
with the twenty four chiral scalars $x^a$.
After imposing the delta function $\delta(p_1-\vec v\cdot
\vec p)$ in the path integral, and performing the 
integral over $p_1$, we now find the partition sum:
\eqn\eee{Z_0={1\over\eta^{24}(\tau)}
\sum_{p\in\Lambda} q^{{1\over2}[p^2-(\vec p\cdot \vec
v)^2]}\bar q^{-{1\over2}(\vec p\cdot \vec v)^2}}
The form in square brackets implies that one combination
of the twenty-four scalars $x^a$ decouples, and we are left with
an effectively 23 dimensional left moving lattice. However,
the $\bar q$ dependence indicates that the missing
combination reappears on the right moving side, with
negative norm. Thus, overall, we find that momenta live
on a 23+1 dimensional lattice where all twenty-four dimensions
have positive norm, the first twenty-three trivially, and the
twenty-fourth because of a cancellation of two minus signs.
One is due to its
right moving nature (as in Narain compactifications); the other
to its being timelike \eee.

Hence, the 2+1 dimensional vacuum leads to a modular
invariant theory; however, the partition sum \eee\ makes it clear
that the vacuum is unstable, due to a tachyonic divergence
arising from states with $\vec p\cdot \vec v\not=0$ (see the
discussion in the text following eq. \dispr).

Note that, as mentioned in the text, a potential
2+2 dimensional vacuum of N=2 strings is rendered
inconsistent by the above analysis. Indeed, if we
try to gauge, \eg,
\eqn\fff{J=\partial x_4+i\partial x_5\ ,}
the constraint $p_4+ip_5=0$ imposed on the
path integral implies that we are summing
over the sublattice of $\Lambda$ with 
$p_4=p_5=0$. This twenty-two dimensional lattice
does not lead to a modular invariant partition
sum, and hence corresponds to an inconsistent theory.
 
\bigskip

\appendix{B}{Comments on quantization.}

While (uncompactified) N=2 heterotic strings
have a field theoretic spectrum, their quantization 
might nevertheless be subtle. We have seen one
aspect of this in this paper: compactifying N=2
heterotic strings on a circle reveals a rich spectrum
of states absent from the target field theoretic description.
A related subtlety arises when one attempts to study the
system at finite temperature by Euclideanizing time and
making it compact.

In addition, it has been pointed out in the past (see \eg\
\ovone) that loop amplitudes in N=2 string theory
appear to differ from those one would write
down in the target space field theory with the same classical
spectrum and interactions. As an example, at one loop,
in addition
to the usual Schwinger parameter,
which corresponds to the imaginary part of the
modulus of the torus $\tau$, one finds in N=2 string theory
a second modulus $u$ related to the U(1) gauge field (see
section {\it 2.1}).  Since this U(1) modulus lives on
the Jacobian of the torus, the measure for integrating
over $u$ involves a factor of ${\it Im}\tau$, which affects
power-counting of momentum integrals \ovone.
Moreover, modular invariance cuts off ultraviolet divergences
in a very non field-theoretic way.

One can in fact see some of these issues classically.
Consider an N=2 string in a linear dilaton background,
$\Phi(x)=Q_\mu x^\mu= Q^i x^{\bar i}+Q^{\bar i} x^i$.
Criticality requires that $Q^2=Q\cdot \bar Q=0$. As is well
known, the linear dilaton modifies the superconformal
generators \ntwo\ to:
\eqn\tmod{\eqalign{
  \bar T=&~-{1\over2}\left(\bar\partial x\cdot 
  	\bar\partial x+Q\cdot \bar\partial^2 x +
	\bar\psi\cdot\bar\partial\bar\psi\right)\cr
  \bar G^\pm=&~(\eta_{\mu\nu}\pm I_{\mu\nu})
	\left(\bar\psi^\mu\bar\partial x^\nu
 	+Q^\mu\bar\partial\psi^\nu\right) \cr
  \bar J=&~\half I_{\mu\nu}\bar\psi^\mu\bar\psi^\nu+
	I_{\mu\nu}Q^\mu\bar\partial x^\nu~.  \cr}
}
The modified form of $\bar J$ leads to a change 
in the physical state conditions.
While before, physical states such as \vbos, \gf\ \etc\
had to satisfy the 
physical state condition $k^2=0$, now one finds 
{\it two} conditions which can be written as:
\eqn\bone{k_\mu(k^\mu+Q^\mu)=0~;\;\;\; k_\mu Q_\nu I^{\mu\nu}=0}
or in complex coordinates:
\eqn\btwo{k\cdot(\bar k+\bar Q)=0~; \;\;\;\bar k\cdot (k+Q)=0.}
The theory behaves as if the kinetic operator itself has been
complexified in going from 1+1 to 2+2d.
Target space field theory in a linear dilaton background
would reproduce naturally the first of equations \bone.
It remains to be understood how it knows about the second.

\listrefs
\end


+Q^\mu)=0~;\;\;\; k_\mu Q_\nu I^{\mu\nu}=0}
or in complex coordinates:
\eqn\btwo{k\cdot(\bar k+\bar Q)=0~; \;\;\;\bar k\cdot (k+Q)=0.}
The theory behaves as if the kinetic operator itself has been
complexified in going from 1+1 to 2+2d.
Target space field theory in a linear dilaton background
would reproduce naturally the first of equations \bone.
It remains to be understood how it knows about the second.

\listrefs
\end